\documentclass[preprint,authoryear,12pt]{elsarticle}

\usepackage{amssymb}

\usepackage{lineno}

\usepackage{color}
\usepackage{amsmath}
\usepackage{subfigure}
\usepackage{graphicx}
\usepackage{graphics}
\usepackage{booktabs}
\usepackage{threeparttable}
\usepackage{appendix}
\usepackage{endfloat}

\journal{}

\begin{document}

\begin{frontmatter}



\title{Route guidance strategies revisited: Comparison and evaluation in an asymmetric two-route traffic system}


\author{Zhengbing He}
\address{MOE Key Laboratory for Urban Transportation Complex Systems Theory and Technology, Beijing Jiaotong University, China. he.zb@hotmail.com}
\author{Ning Jia}
\address{Institute of Systems Engineering, Tianjin University, China. jianing.blgt@gmail.com}
\author{Wei Guan (corresponding author)}
\address{MOE Key Laboratory for Urban Transportation Complex Systems Theory and Technology, Beijing Jiaotong University, China. weig@bjtu.edu.cn}

\begin{abstract}
To alleviate traffic congestion, a variety of route guidance strategies has been proposed for intelligent transportation systems.
A number of the strategies are proposed and investigated on a symmetric two-route traffic system over the past decade.
To evaluate the strategies in a more general scenario, this paper conducts eight prevalent strategies
on a asymmetric two-route traffic network with different slowdown behaviors on alternative routes.
The results show that only mean velocity feedback strategy is able to equalize travel time, i.e., approximate user optimality;
while the others fail due to incapability of establishing relations between the feedback parameters and travel time.
The paper helps better understand these strategies, and suggests mean velocity feedback strategy if the authority intends to achieve user optimality.

\end{abstract}


\end{frontmatter}

\linenumbers


\newpage

\section{Introduction}
Nowadays traffic congestion has been one of the most prevalent city diseases.
To alleviate the congestion, route guidance strategies, which recommend optimal route for drivers, are receiving extensive attention
\citep[see e.g.][]{Pavlis1999a,Deflorio2000, Wahle2002, Berg2004, Hegyi2004, Liu2011, Nagatani2011,Tobita2012}.
Over the past decade, a variety of route guidance strategies has also been proposed and investigated in the field of physics, such as
travel time feedback strategy (TTFS) \citep{Wahle2000},
mean velocity feedback strategy (MVFS) \citep{Lee2001},
congestion coefficient feedback strategy (CCFS) \citep{Wang2005a},
prediction feedback strategy (PFS) \citep{Dong2009},
weighted congestion coefficient feedback strategy (WCCFS) \citep{Dong2010a},
corresponding angle feedback strategy (CAFS) \citep{Dong2010c},
vehicle number feedback strategy (VNFS) \citep{Dong2010b},
vacancy length feedback strategy (VLFS) \citep{Chen2012}, etc.
All these strategies are proposed and studied in a symmetric two-route traffic network first adopted in \citet{Wahle2000}.
The remarkable features of the traffic system are not only the same configurations of alternative routes,  
but also the same slowdown probability utilized in the employed Nagel-Schrekenberg model, which reflects drivers' imperfect break behaviors.
However, the slowdown probability pertaining to routes is not the same most of time in reality
due to different traffic conditions on alternative routes, such as different road types, different percentages of trucks, etc.

The paper is thus dedicated to evaluating the effectiveness of the existing strategies in a more general asymmetric two-route network with different slowdown probability on alternative routes.
The research will help better understand these strategies and provide implications for practical applications of the strategies.
Toward the end, the remainder of the paper is organized as follows:
section \ref{sec:GuidanceStrategy} briefly describes eight route guidance strategies revisited in the paper;
section \ref{sec:NSModelAndNetwork} introduces the traffic flow model and the user optimality;
section \ref{sec:results} compares and evaluates the performance of these strategies in symmetric and asymmetric traffic systems;
conclusions are made at last.

\section{The existing route guidance strategies}\label{sec:GuidanceStrategy}
The paper revisits the following eight route guidance strategies:

$\bullet$ TTFS diverts an incoming vehicle to a route with the minimum travel time of the last departure vehicle.
When a route is empty, the travel time is set as free flow travel time, i.e., dividing route length by maximum velocity.

$\bullet$ MVFS diverts an incoming vehicle to a route with the minimum mean velocity of all en-route vehicles.

$\bullet$ CCFS diverts an incoming vehicle to a route with the minimum congestion coefficient, which is defined as

\begin{equation}
    C= \sum_{i=1}^m (n_i)^w
\end{equation}
where $n_i$ is vehicle number of the \emph{i}th congestion cluster in which vehicles are close to each other without any gap; $w$ is a weight coefficient, and the value is set as 2 as \cite{Wang2005a} did; $m$ is the total number of en-route vehicles.

$\bullet$ PFS diverts an incoming vehicle to a route with the minimum congestion coefficient in prediction time. Prediction is made in the same simulation scenario, and the value is set as 60s as \cite{Dong2009} suggested.

$\bullet$ WCCFS diverts an incoming vehicle to a route with the minimum weight congestion coefficient, which is proposed as

\begin{equation}
    C_w= \sum_{i=1}^m F(l_i)(n_i)^w= \sum_{i=1}^m \left(k\frac{l_i}{2000}+2\right)(n_i)^w
\end{equation}
where $F(l_i)$ is a weight function; $l_i$ is the location of the middle vehicle in the \emph{i}th congestion cluster; $k$ and $w$ are respectively set as -2 and 2 as suggested in \cite{Dong2010a}.

$\bullet$ CAFS diverts an incoming vehicle to a route with the minimum corresponding angle congestion coefficient, which reads

\begin{equation}
    C_\theta= \sum_{i=1}^m \theta_i^2 = \sum_{i=1}^m \left( \arctan \left(\frac{n_i^{first}}{H}\right) - \arctan \left(\frac{n_i^{last}-1}{H}\right)  \right)^2
\end{equation}
where $\theta_i$ stands for a corresponding angle of the \emph{i}th congestion cluster; $n_i^{first}$ and $n_i^{last}$ are the locations of the first and last vehicles in the \emph{i}th congestion cluster; $H$ is set as 100 as \cite{Dong2010c} did.

$\bullet$ VNFS diverts an incoming vehicle to a route with the minimum number of en-route vehicles.

$\bullet$ VLFS diverts an incoming vehicle to a route with the maximum distance between entrance and the vehicle closest to the entrance.

Obviously, CCFS, PFS, WCCFS and CAFS are based on measuring the congestion cluster; we thus call them as cluster-based strategies.

\section{The traffic flow model and the user optimality}\label{sec:NSModelAndNetwork}
As the previous works proposing the strategies did, the paper also employs the Nagel-Schrekenberg (NS) model \citep{Nagel1992},
one of the most popular traffic flow models presently.
Due to popularity of the NS model, we do not introduce the details except for the slowdown probability
(for more information one can refer to relevant papers such as those cited in the introduction).
The slowdown probability incorporated in the NS model is to reflect drivers' imperfect break behaviors,
i.e., at each simulation step a vehicle reduces its velocity by one unit with the slowdown probability.
With consideration of the probability, macroscopic traffic phenomena such as stop-and-go waves, traffic breakdown, can be well reproduced by the NS model.
Obviously, the larger the probability is, the larger the number of congestion clusters is.


The diversion is at each time step generating a new vehicle that chooses a route completely following the recommendation of the route guidance strategies;
the new vehicle will be deleted if its target loading cell (the first cell) is occupied.
Note that drivers' completely follow implies that all outcomes are led by the strategies; it is benefit to evaluate the strategies.

Compared with optimizing traffic conditions of a route guidance strategy,
we are more interested in its ability of achieving user optimality (UO),
i.e., equalizing travel times on two routes.
It is because that UO is compatible with drivers' own decision criteria,
whereas system optimality would sacrifice some drivers' benefits;
it may lead to large scale rejection of route guidance in the long practical run
\citep{Chiu2007,Papageorgiou2007}.

\section{Simulation results}\label{sec:results}
\subsection{A scenario on a symmetric two-route traffic network}
To be convenient for comparison, we first test the strategies on a symmetric traffic system as the previous works did,
i.e., a network with two alternative routes, one entrance and two exits.
More specifically, the routes have the same length 2000 cells, the same maximum velocity 3 cells and the same slowdown probability $p_1=p_2=0.25$.
10000 time steps are conducted and only the stable results from 5000 to 10000 are presented here.
Moreover, we take experienced travel time of departure vehicles as travel time on a route.

\begin{figure}
    \centering
    \subfigure[TTFS]{
    \includegraphics[width=2in]{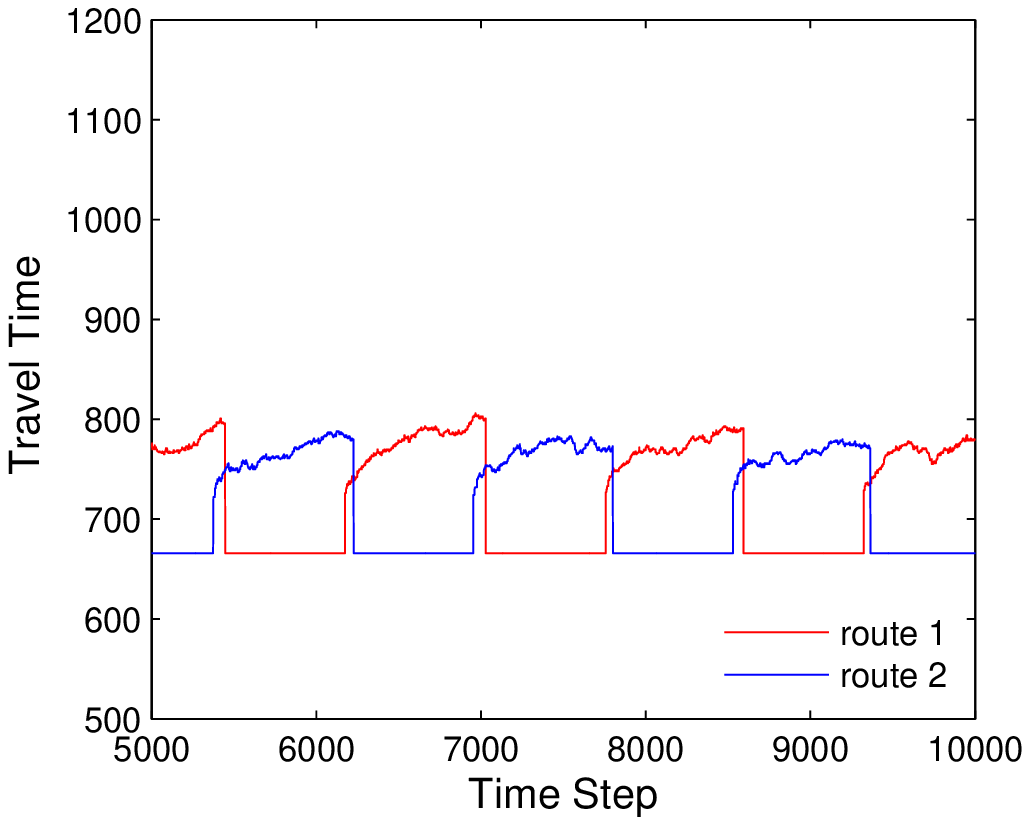}}
    \subfigure[MVFS]{
    \includegraphics[width=2in]{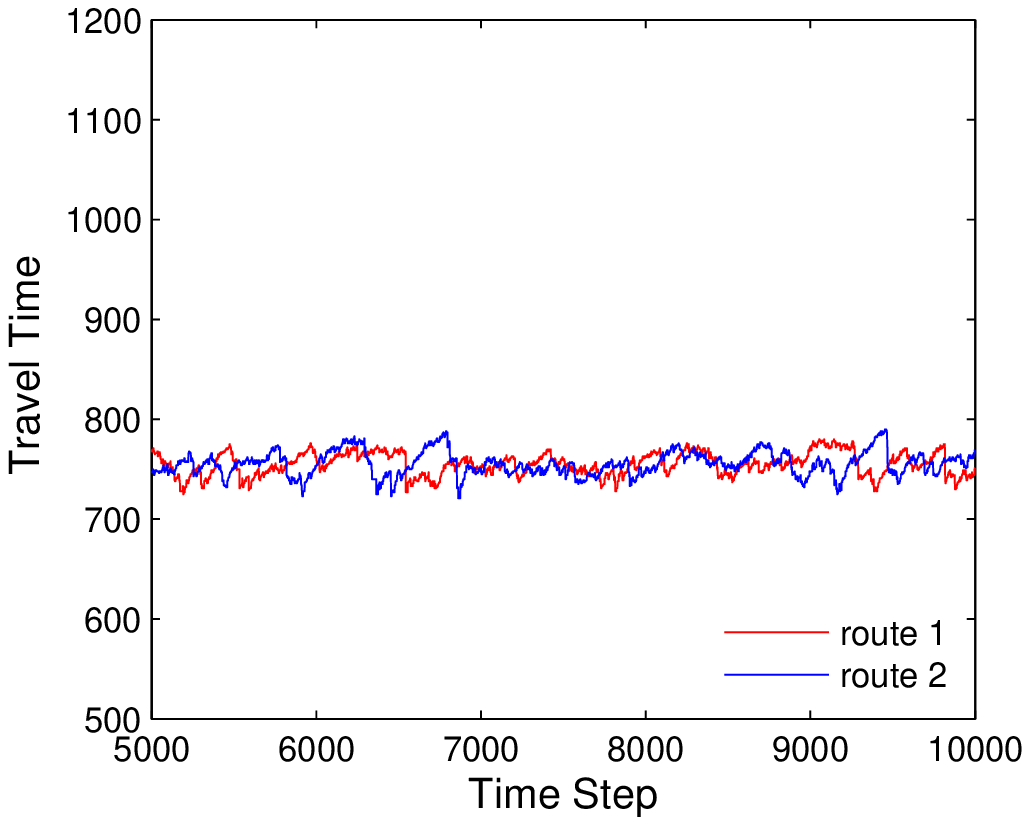}}
    \subfigure[CCFS]{
    \includegraphics[width=2in]{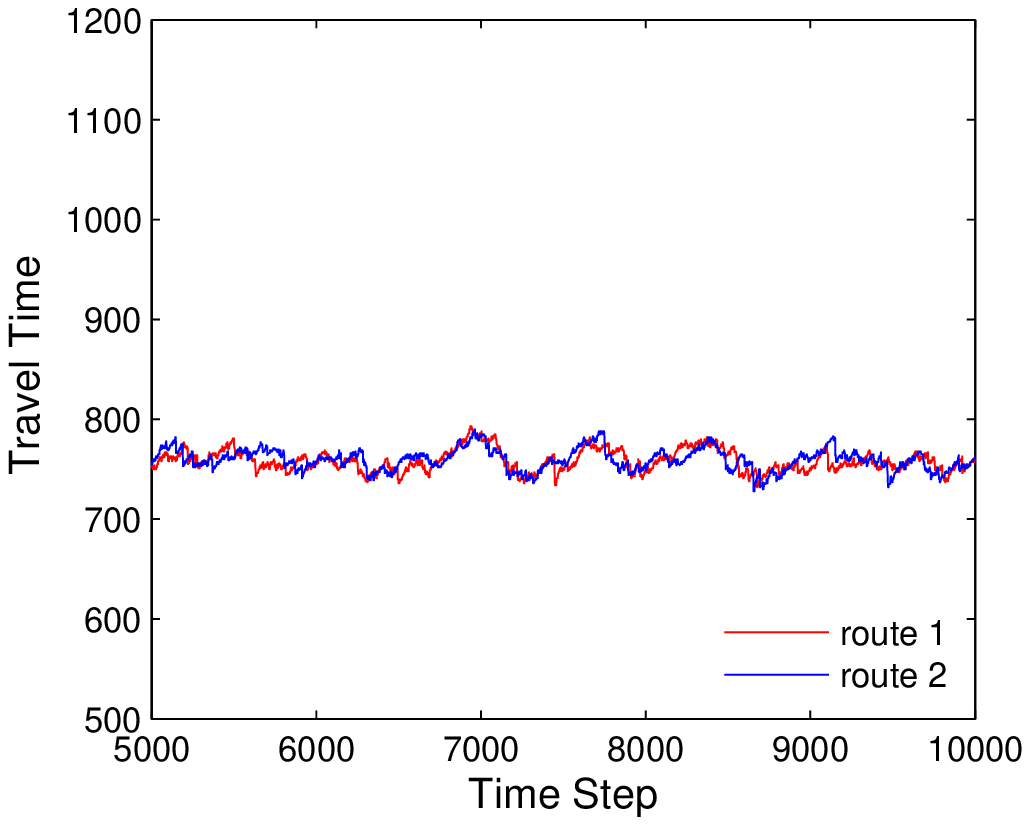}}
    \subfigure[PFS]{
    \includegraphics[width=2in]{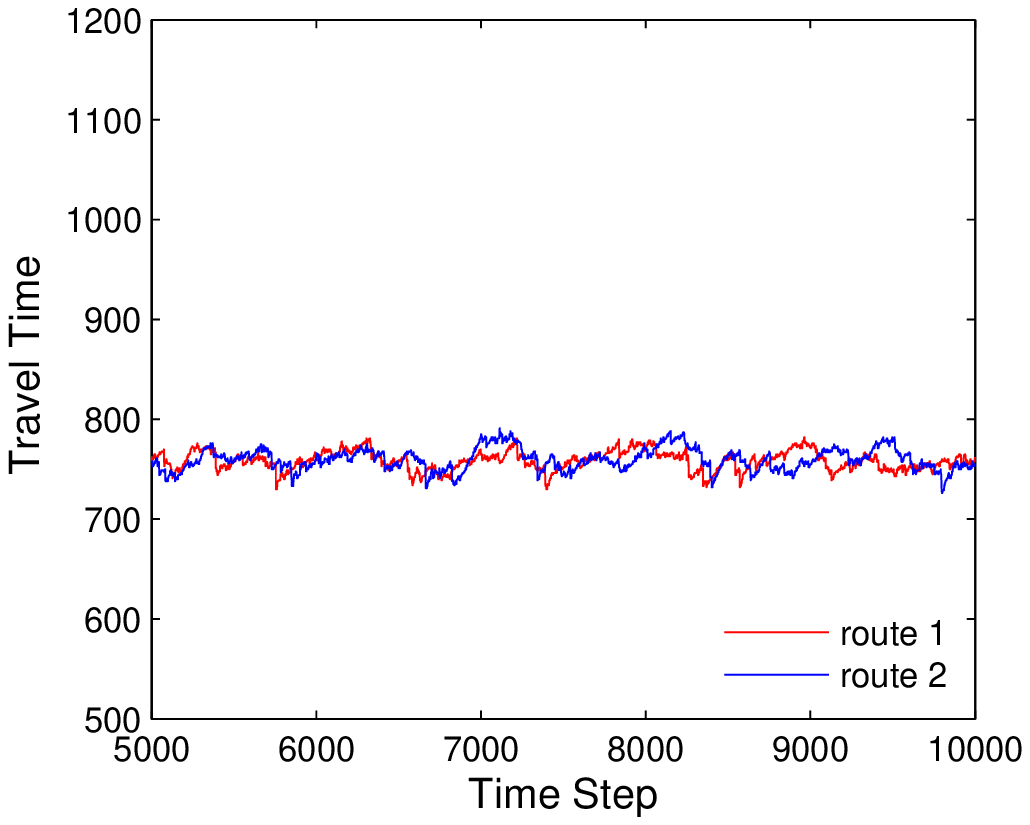}}
    \subfigure[WCCFS]{
    \includegraphics[width=2in]{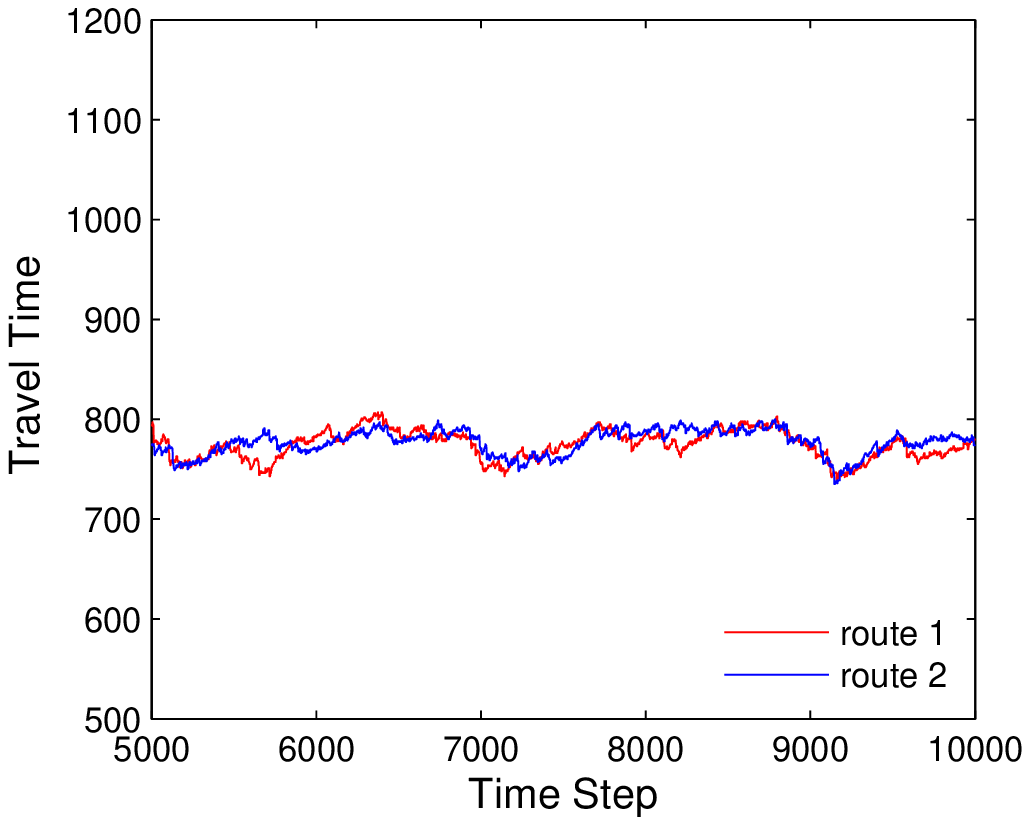}}
    \subfigure[CAFS]{
    \includegraphics[width=2in]{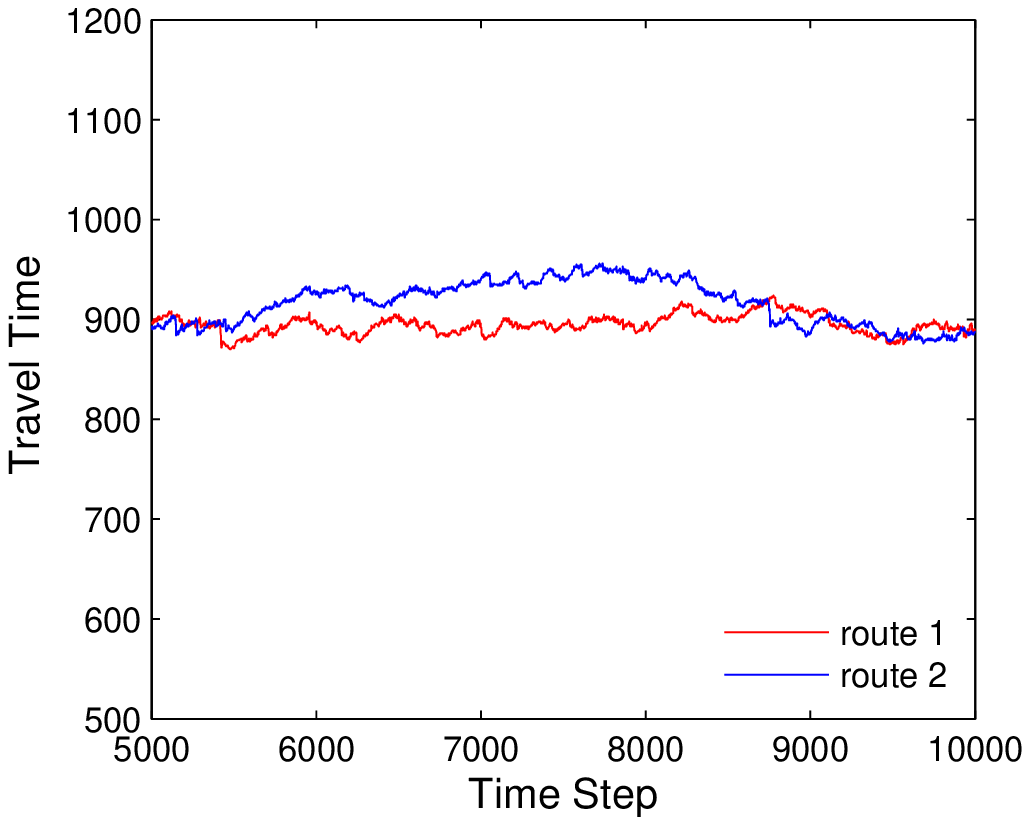}}
    \subfigure[VNFS]{
    \includegraphics[width=2in]{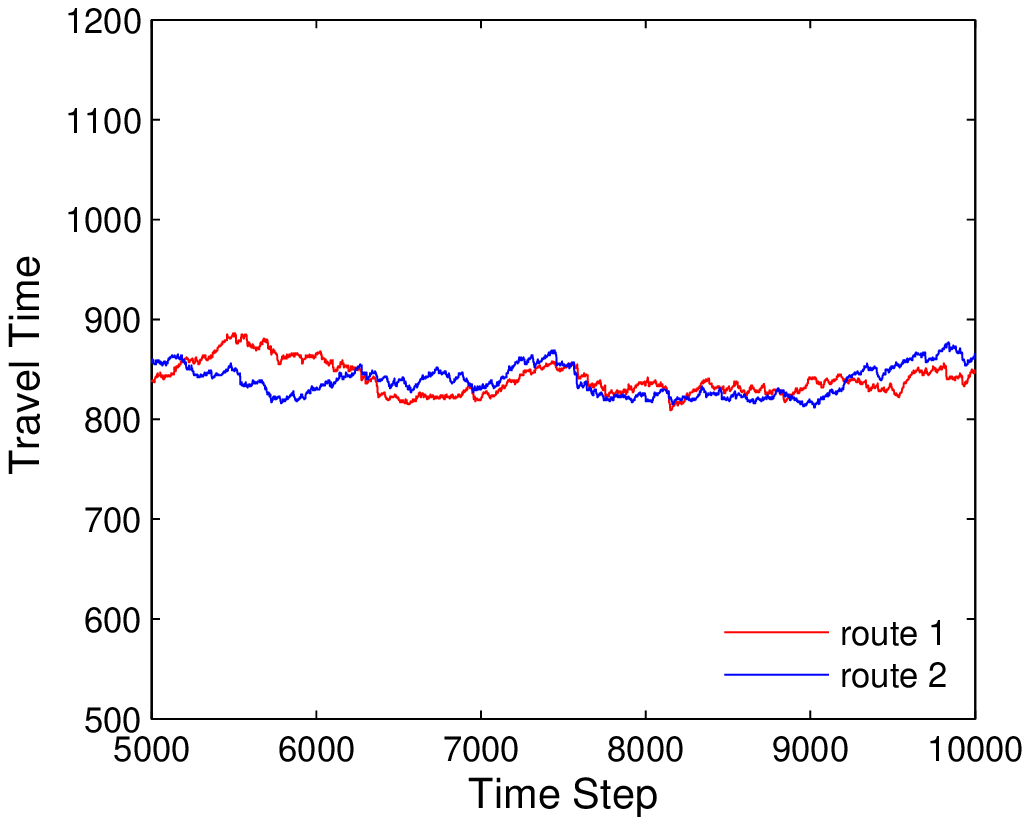}}
    \subfigure[VLFS]{
    \includegraphics[width=2in]{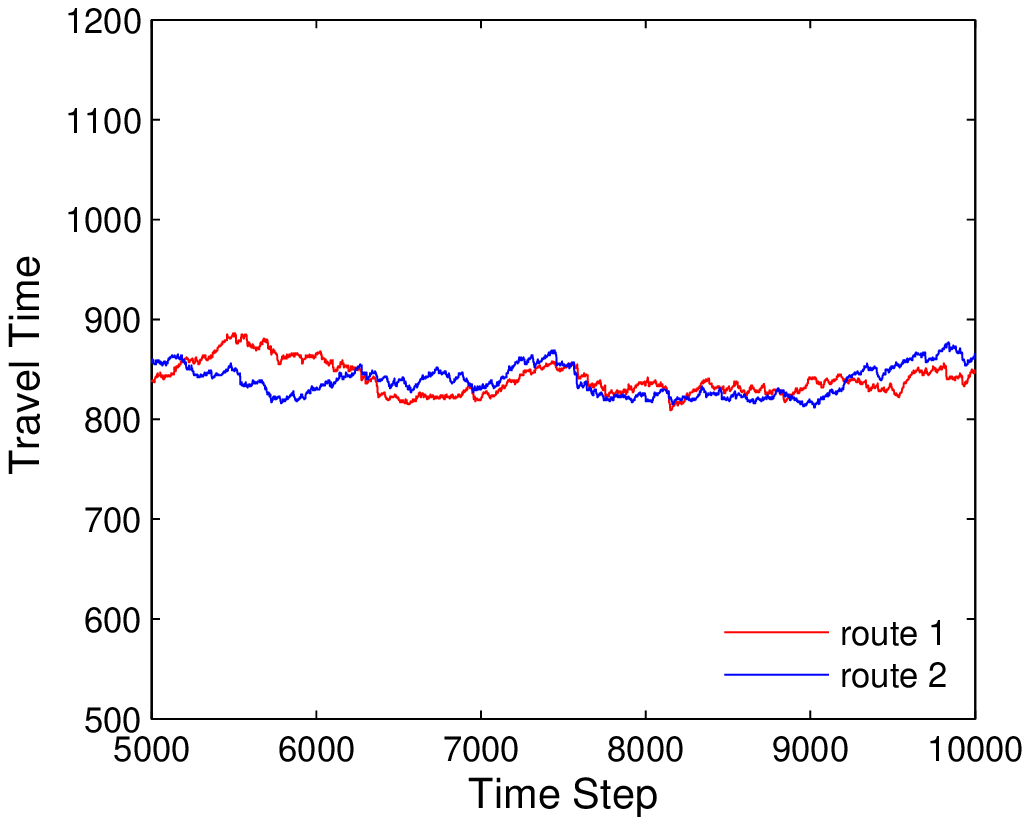}}
    \caption{Travel time on the symmetric network.}
    \label{fig:SymmetricTravelTime}
\end{figure}

\begin{figure}
    \centering
    \subfigure[TTFS]{
    \includegraphics[width=2in]{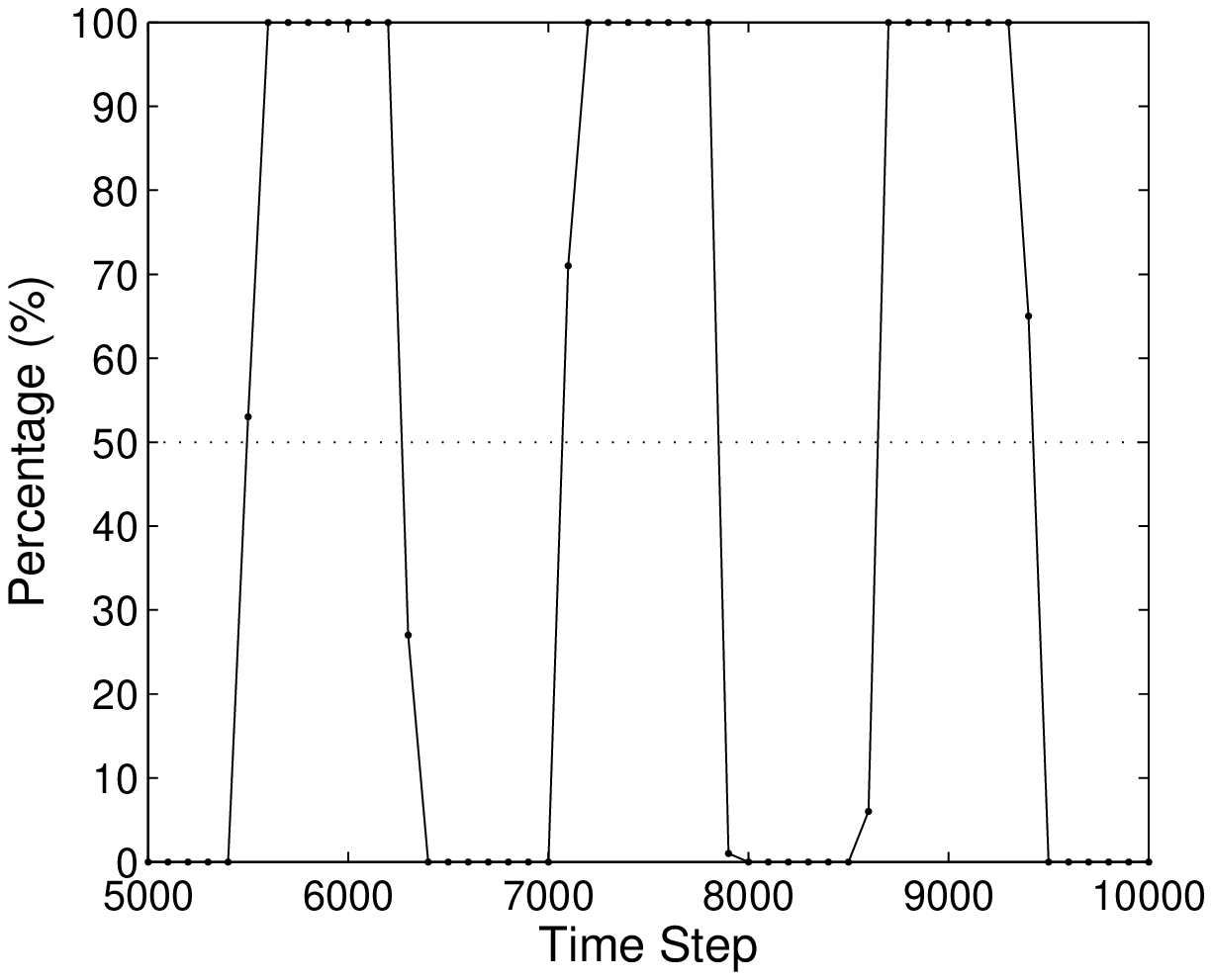}}
    \subfigure[MVFS]{
    \includegraphics[width=2in]{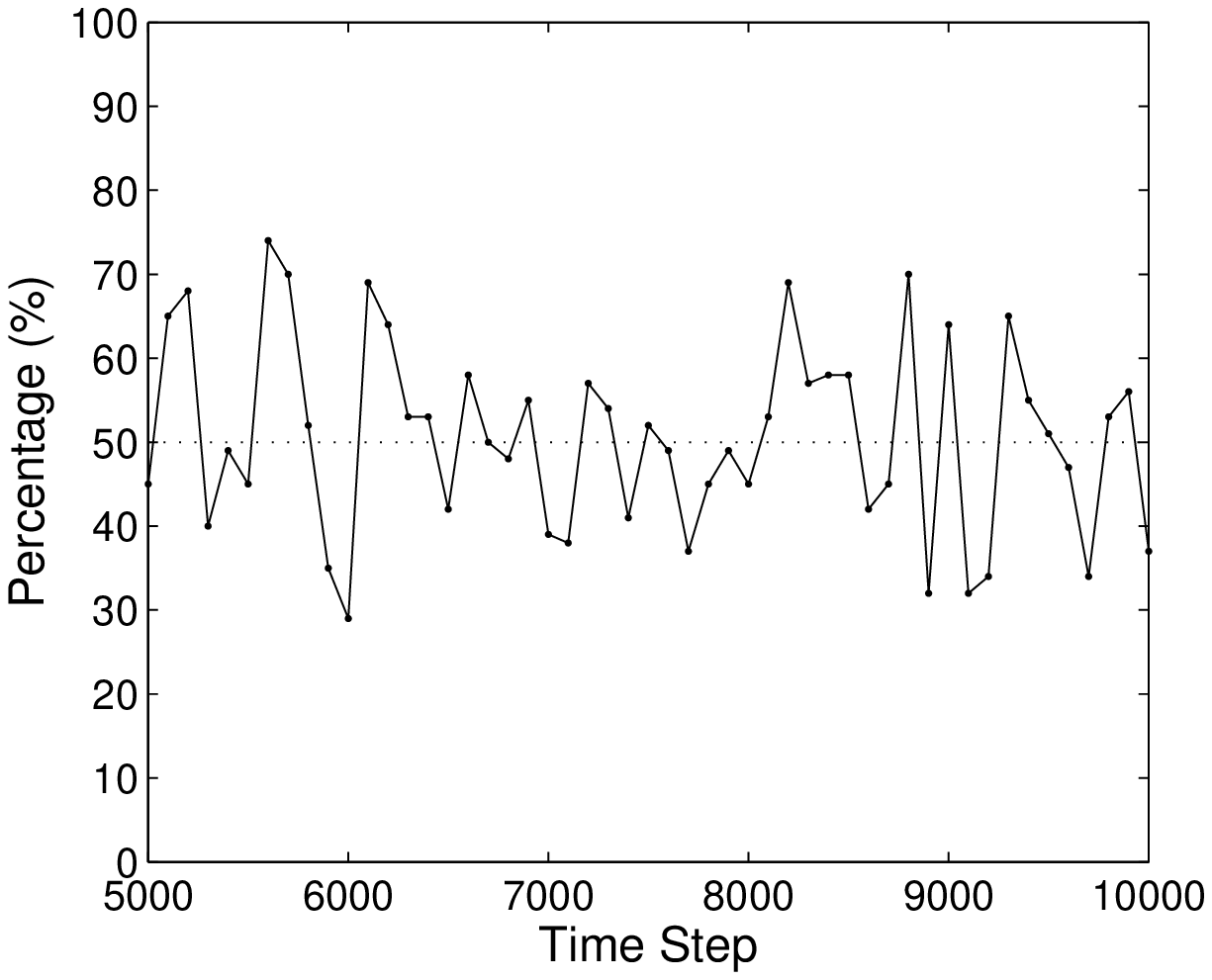}}
    \subfigure[CCFS]{
    \includegraphics[width=2in]{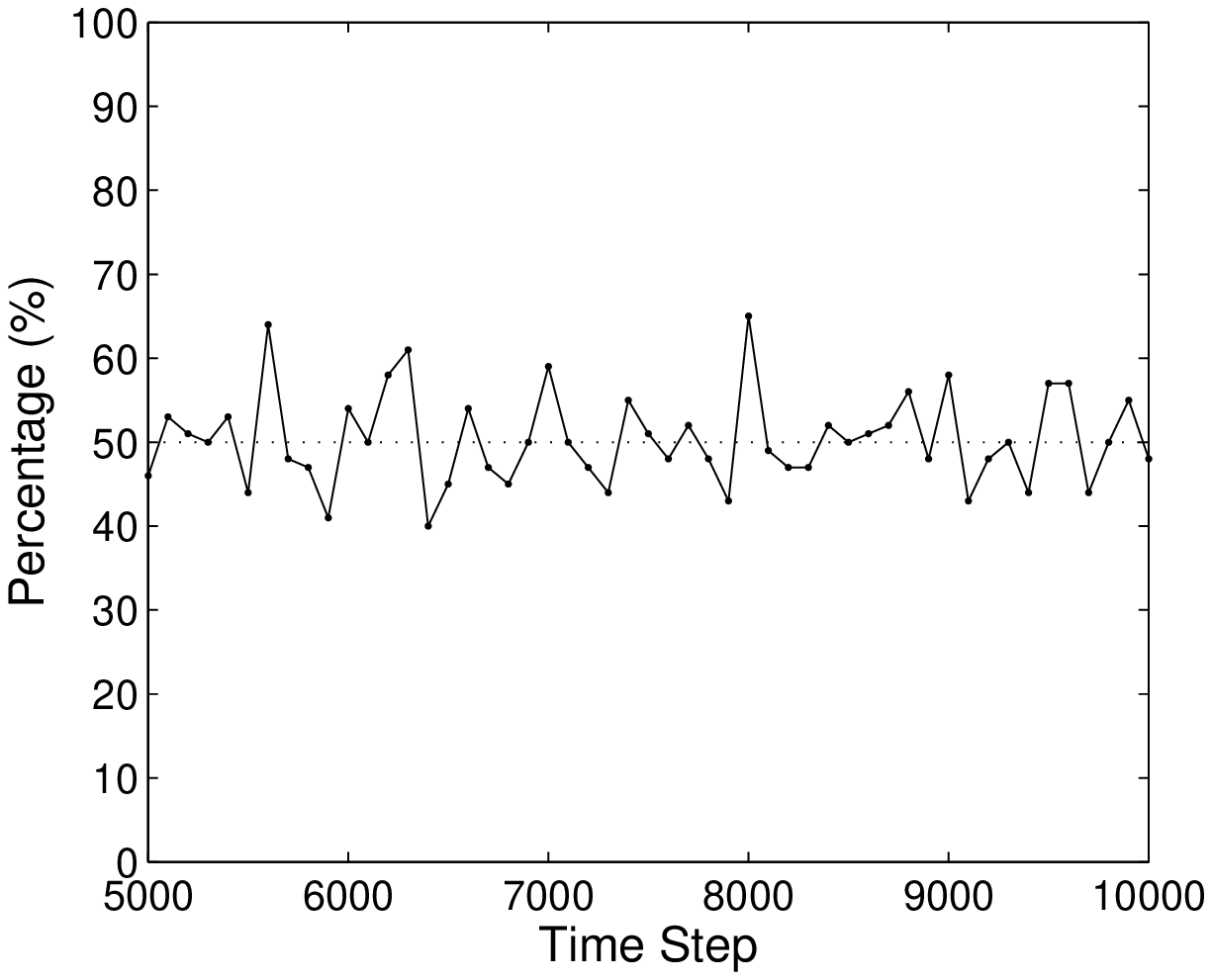}}
    \subfigure[PFS]{
    \includegraphics[width=2in]{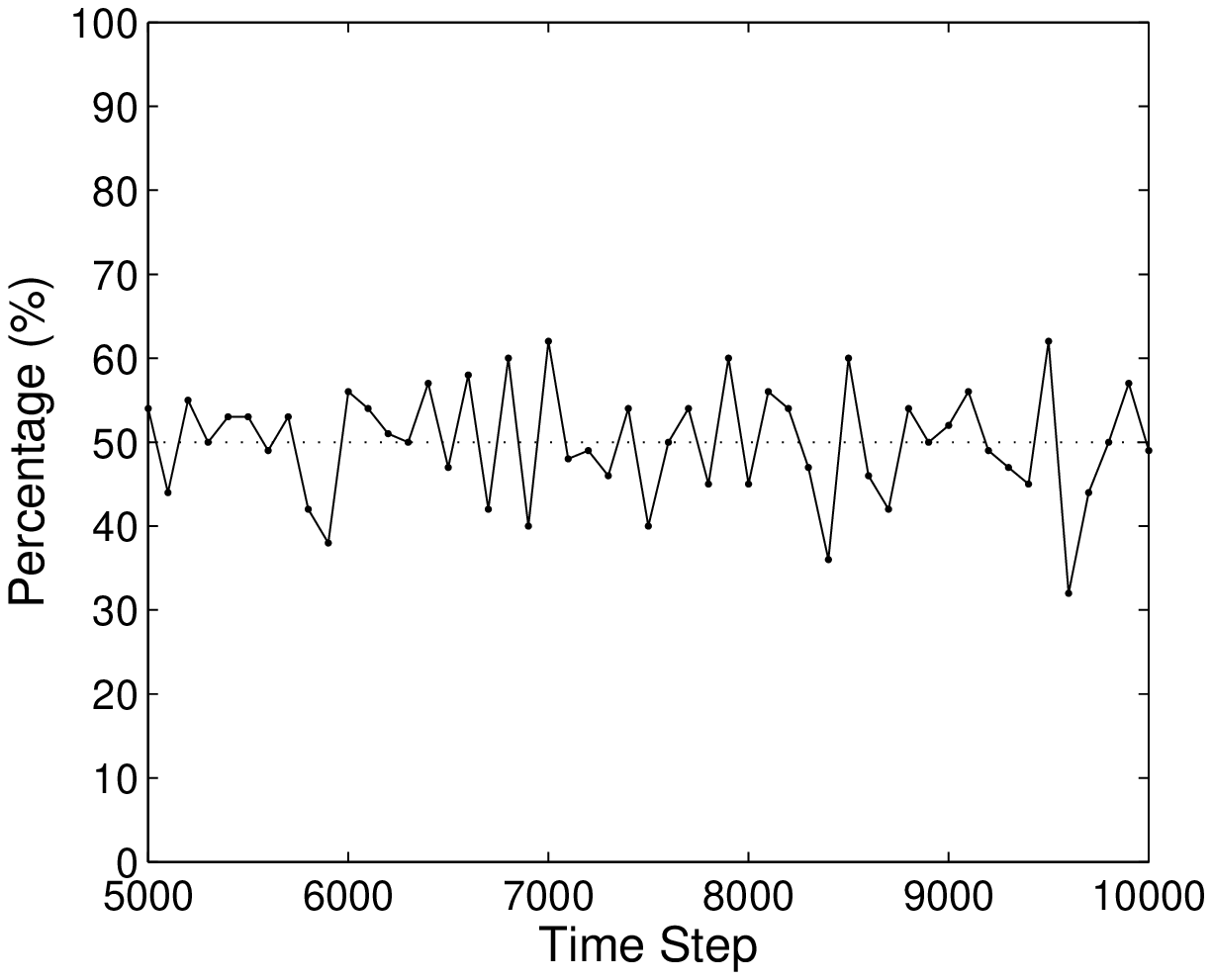}}
    \subfigure[WCCFS]{
    \includegraphics[width=2in]{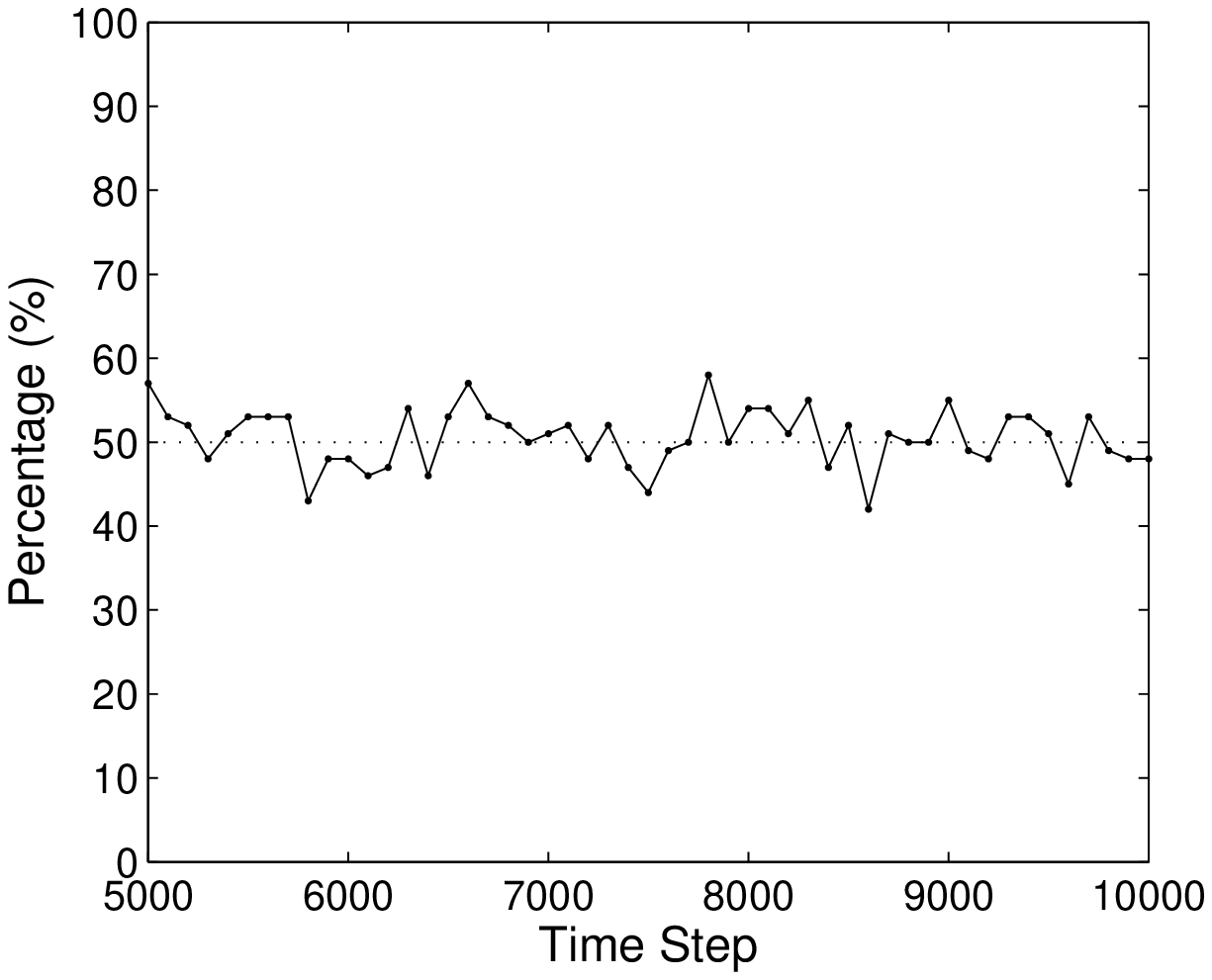}}
    \subfigure[CAFS]{
    \includegraphics[width=2in]{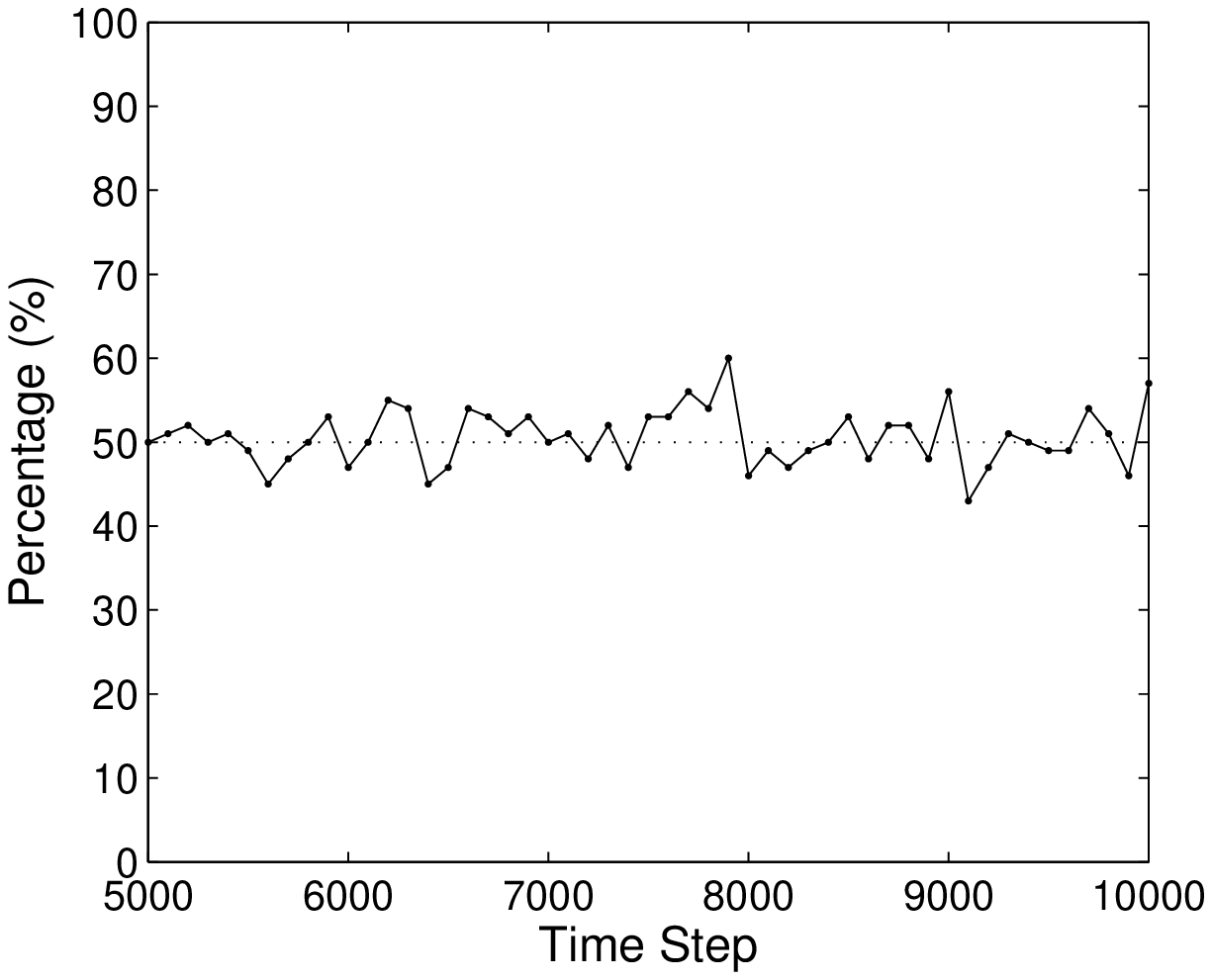}}
    \subfigure[VNFS]{
    \includegraphics[width=2in]{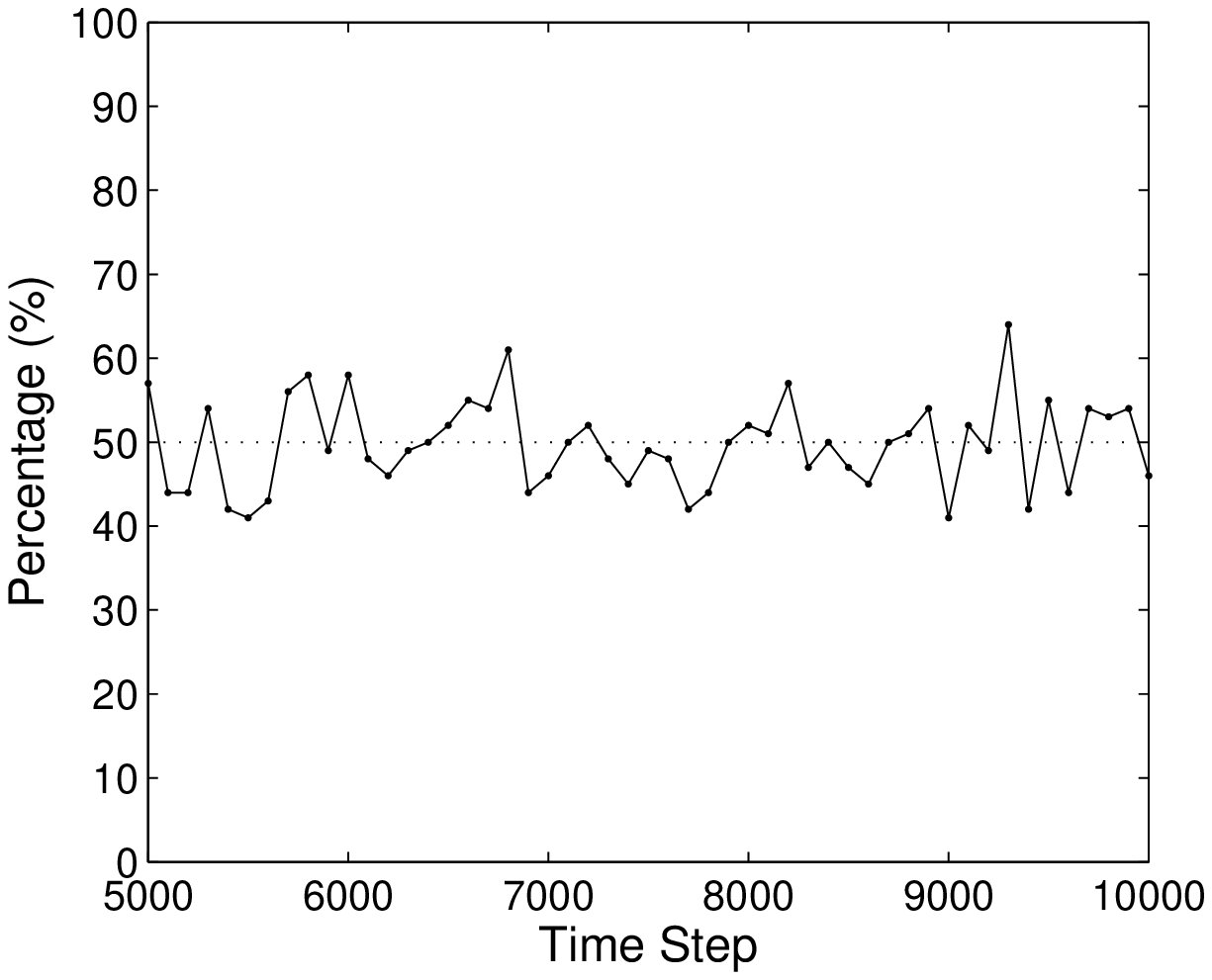}}
    \subfigure[VLFS]{
    \includegraphics[width=2in]{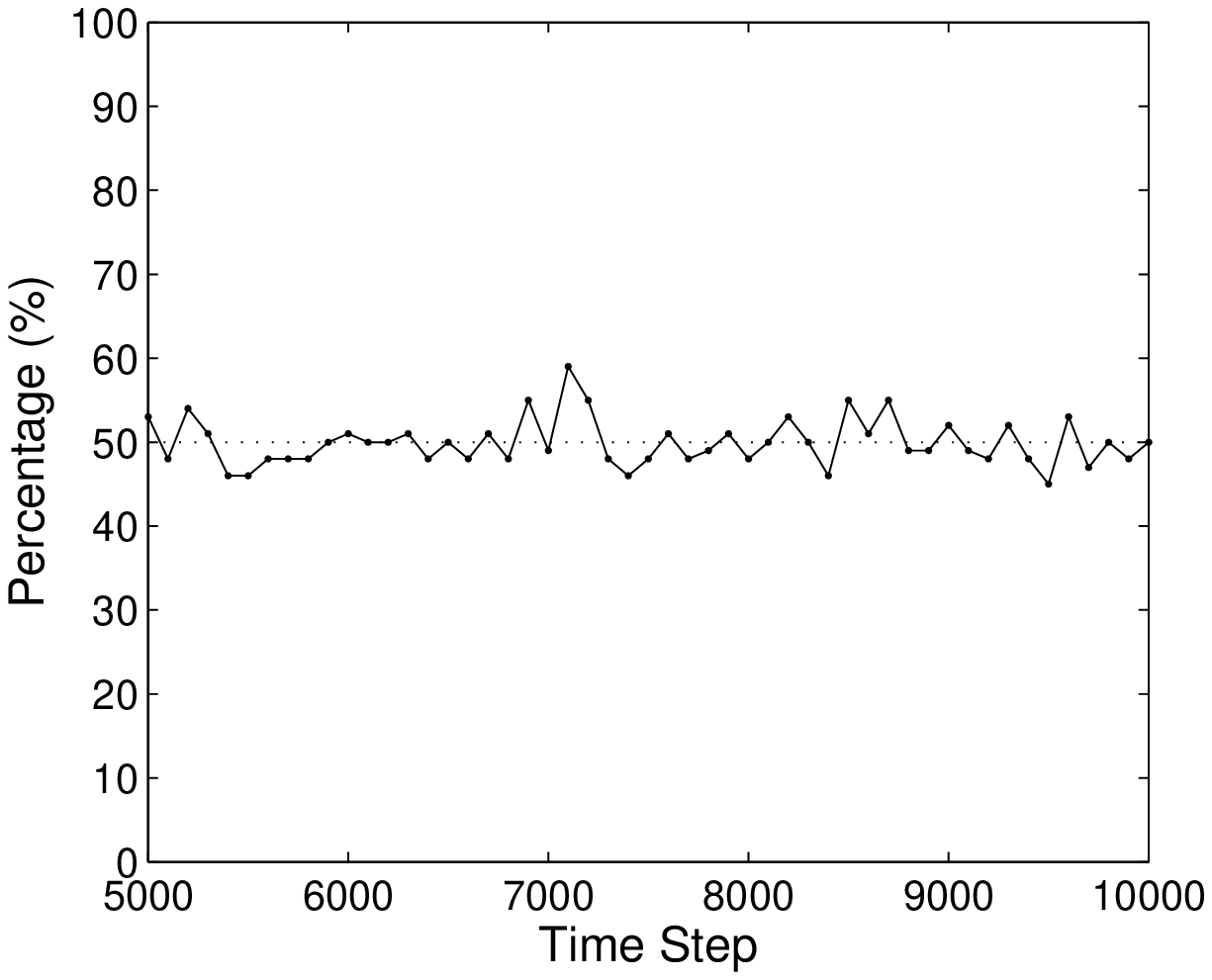}}
    \caption[Vehicle percentages diverting on route 1 on the symmetric network]
            {Vehicle percentages diverting on route 1 on the symmetric network
            (each rate is measured within 100 time steps).}
    \label{fig:SymmetricSplitRate}
\end{figure}

Figure \ref{fig:SymmetricTravelTime} presents the simulation results.
It is obvious that all strategies except for TTFS approximate UO in the scenario.
Figure \ref{fig:SymmetricSplitRate} show vehicle percentages diverting to route 2;
all strategies except for TTFS result in around 50\% diversions due to identity of the routes.
The failure of TTFS can be explained by the lag effect as the previous works mentioned,
i.e., the experienced traffic conditions of a departure vehicle can not reflect the conditions behind it.

\subsection{A scenario on an asymmetric two-route traffic network}
In the section, we apply the existing strategies on a more general asymmetric two-route traffic network, i.e.,
the alternative routes have different slowdown probability, say $p_1=0.4$ and $p_2=0.2$,
the traffic conditions on route 1 are worse than those on route 2; the others are unchanged.

\begin{figure}
    \centering
    \subfigure[TTFS]{
    \includegraphics[width=2in]{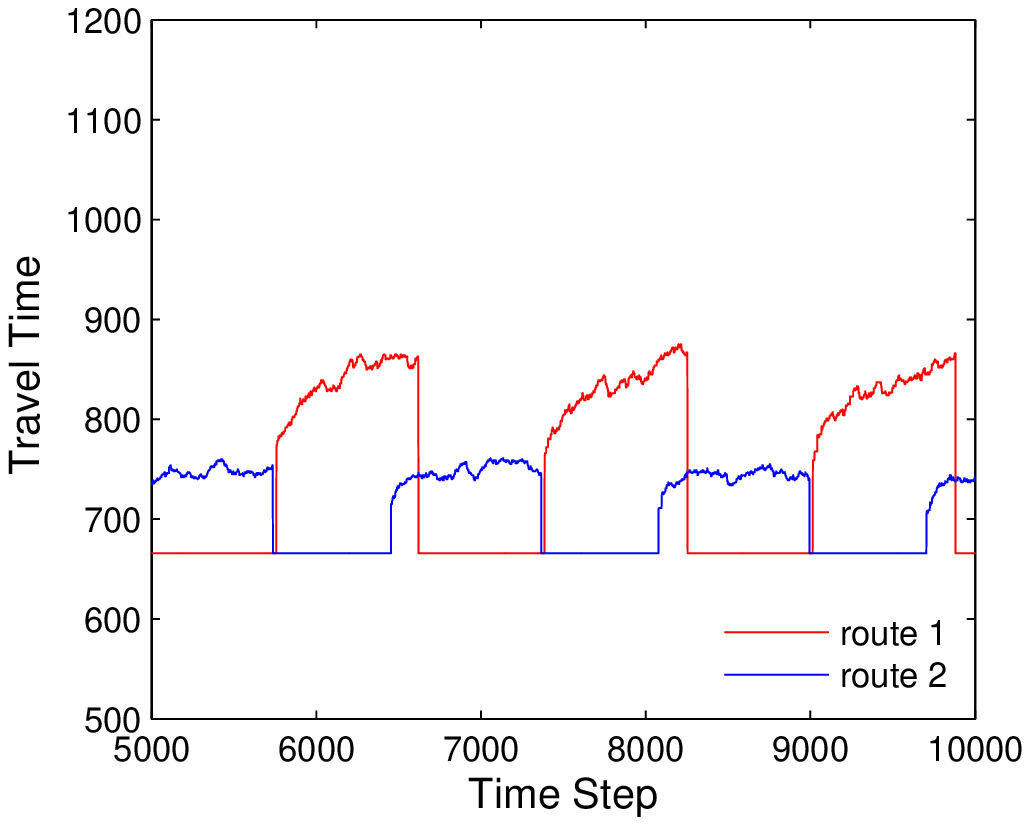}}
    \subfigure[MVFS]{
    \includegraphics[width=2in]{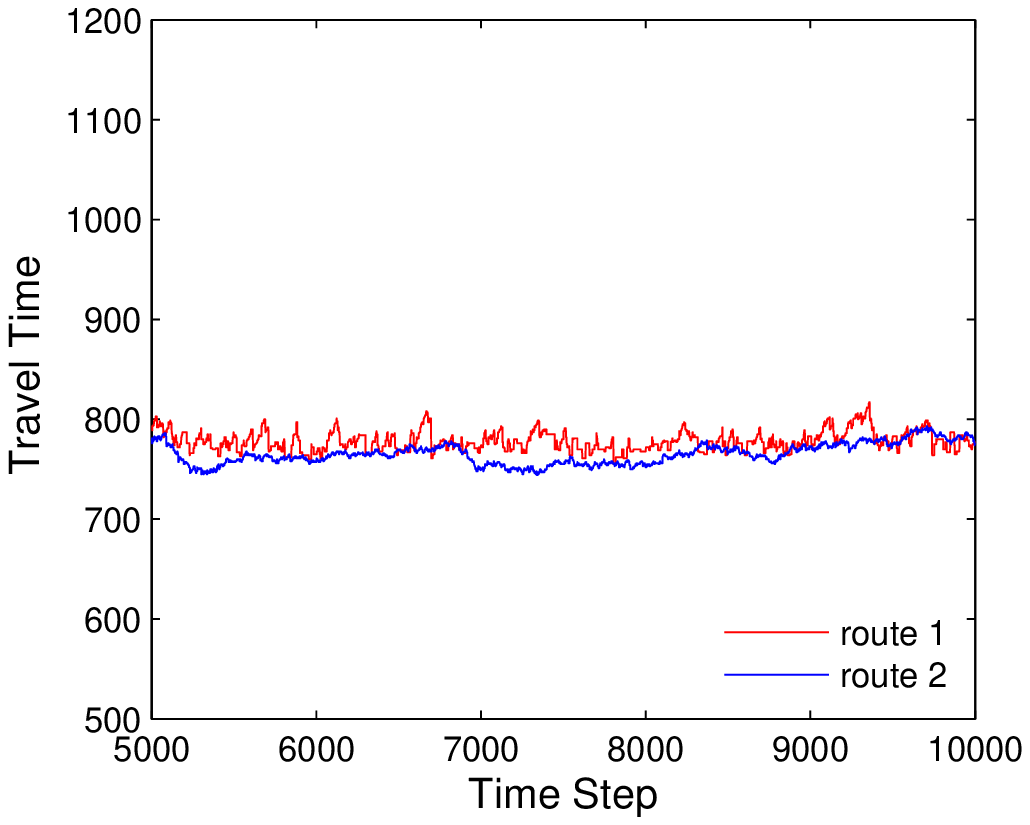}}
    \subfigure[CCFS]{
    \includegraphics[width=2in]{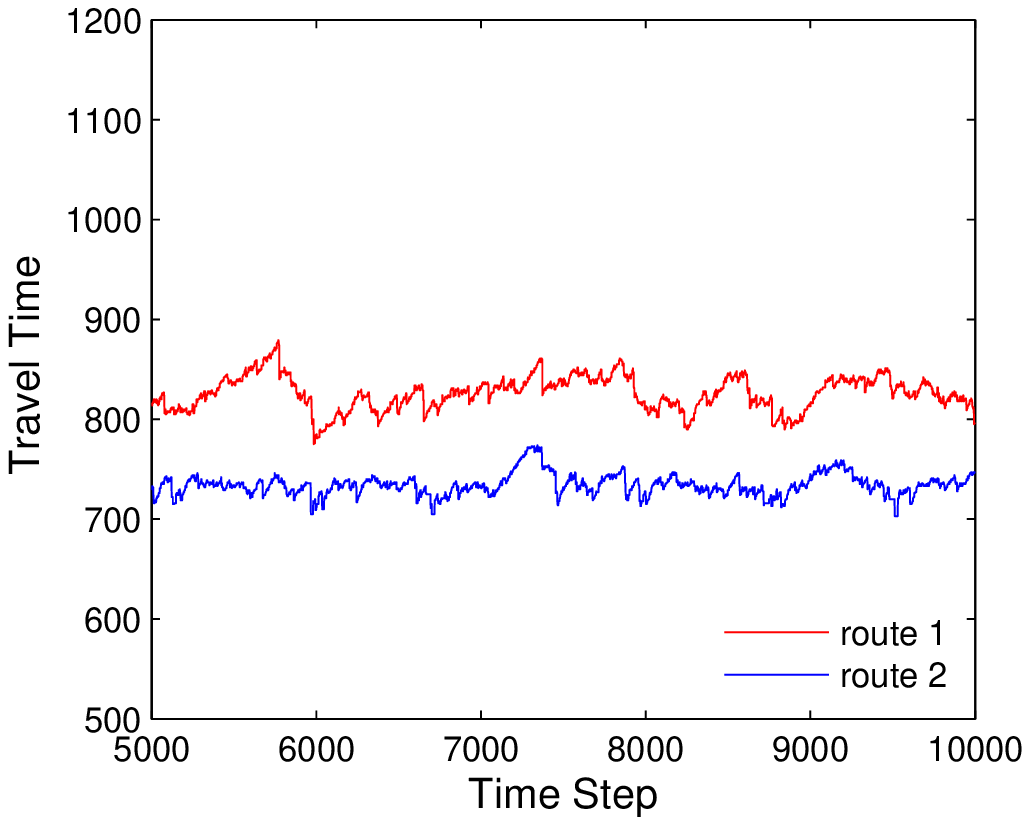}}
    \subfigure[PFS]{
    \includegraphics[width=2in]{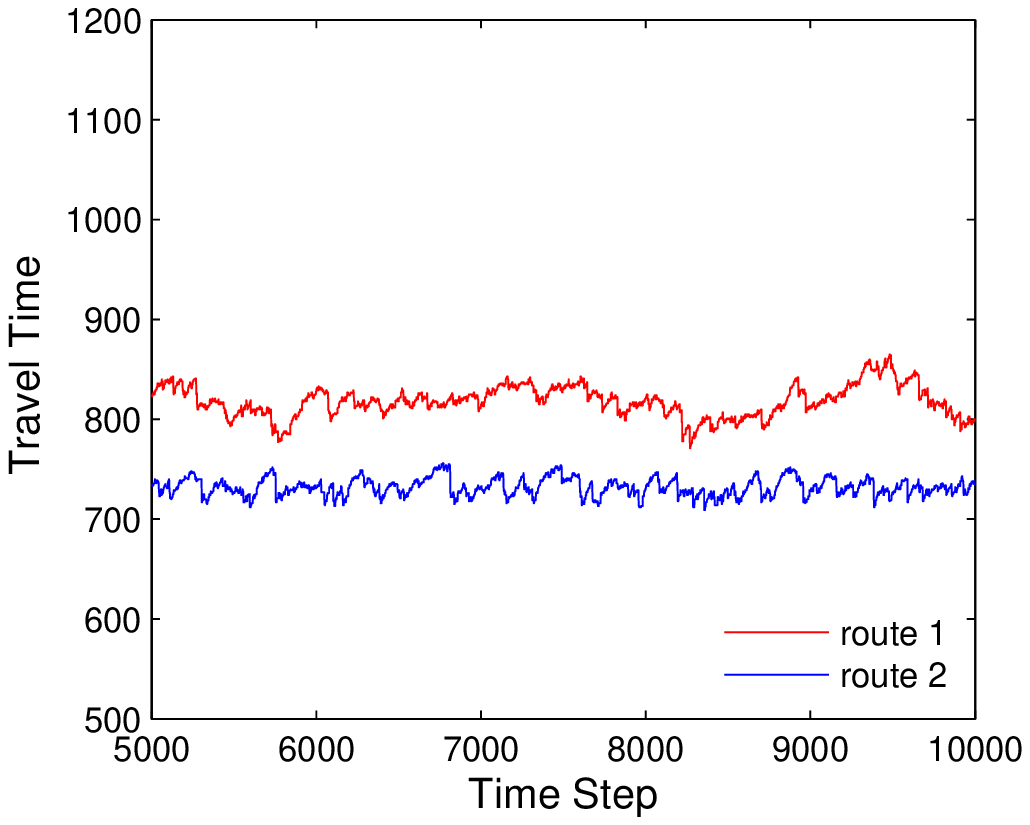}}
    \subfigure[WCCFS]{
    \includegraphics[width=2in]{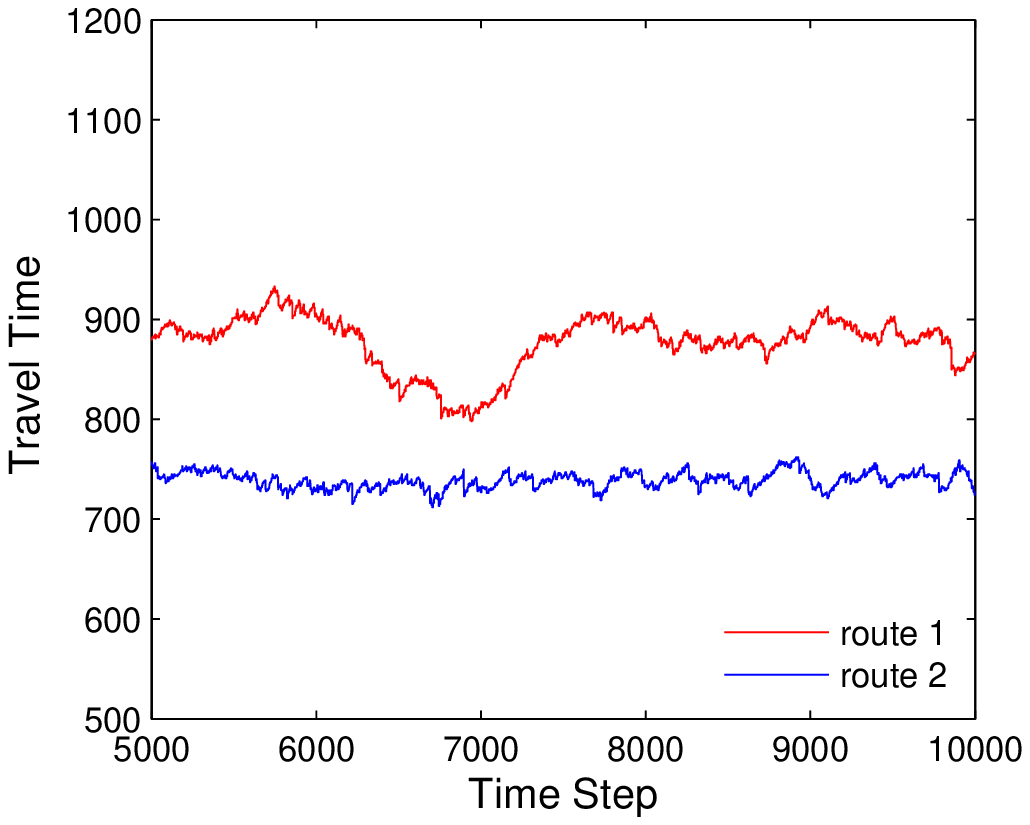}}
    \subfigure[CAFS]{
    \includegraphics[width=2in]{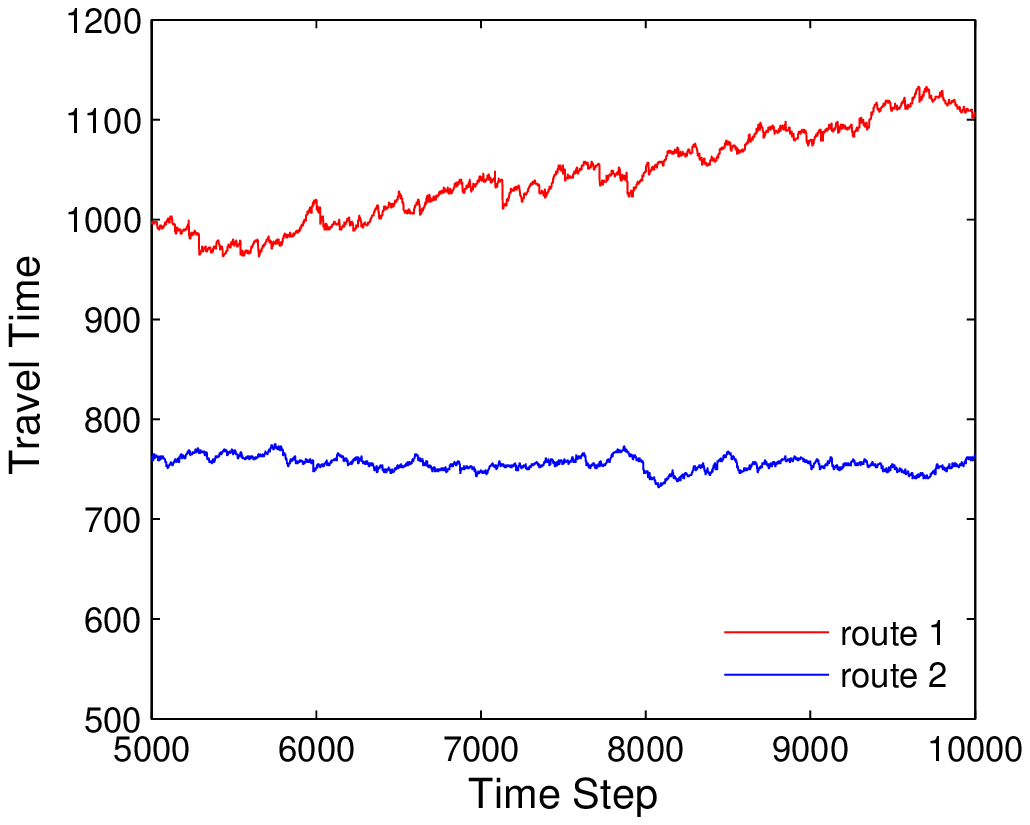}}
    \subfigure[VNFS]{
    \includegraphics[width=2in]{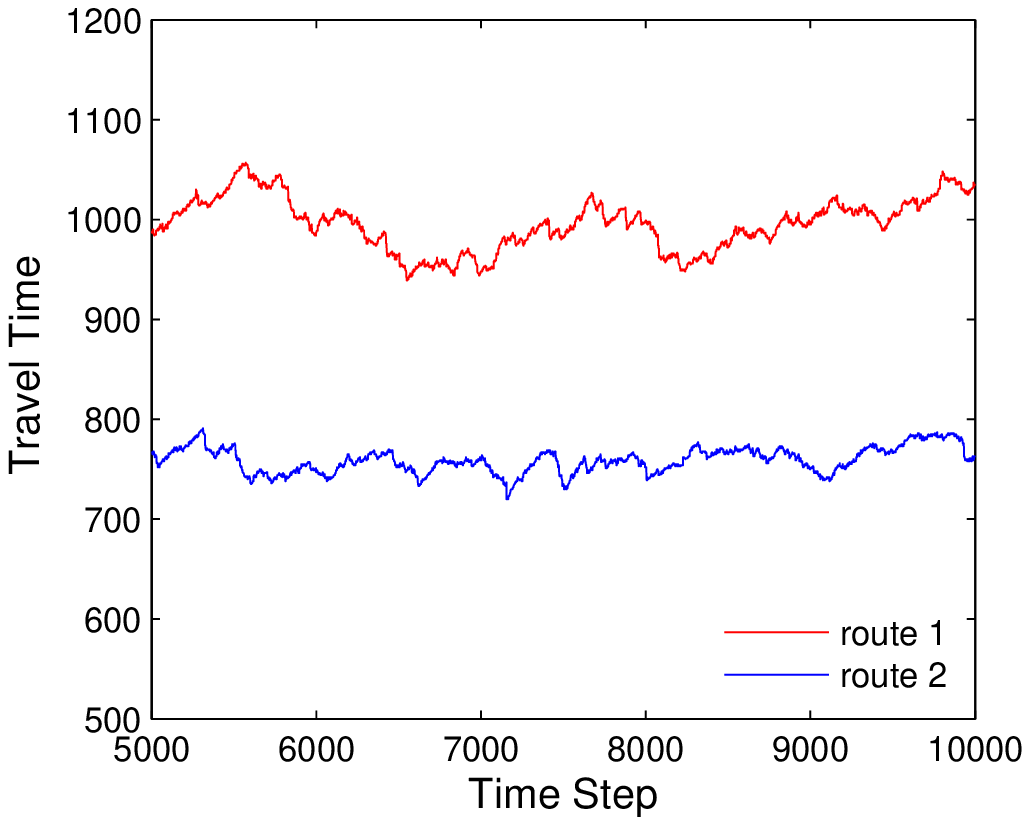}}
    \subfigure[VLFS]{
    \includegraphics[width=2in]{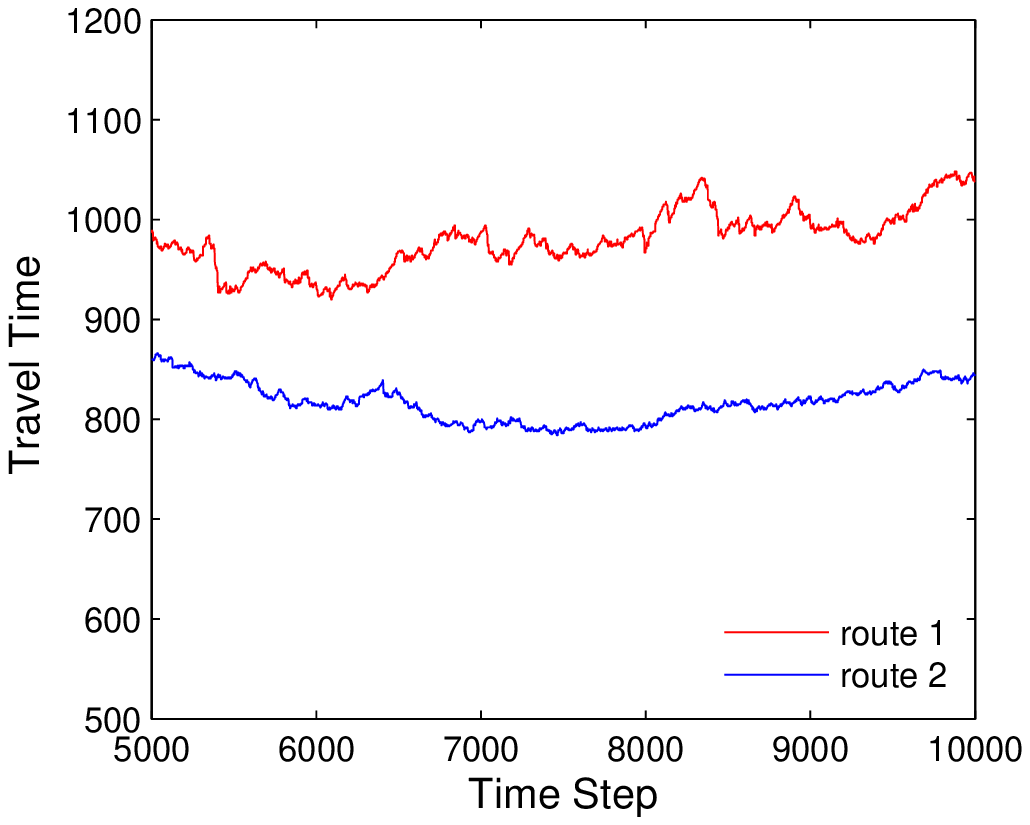}}
    \caption{Travel time on the asymmetric network.}
    \label{fig:TravelTime}
\end{figure}

\begin{figure}
    \centering
    \subfigure[TTFS]{
    \includegraphics[width=2in]{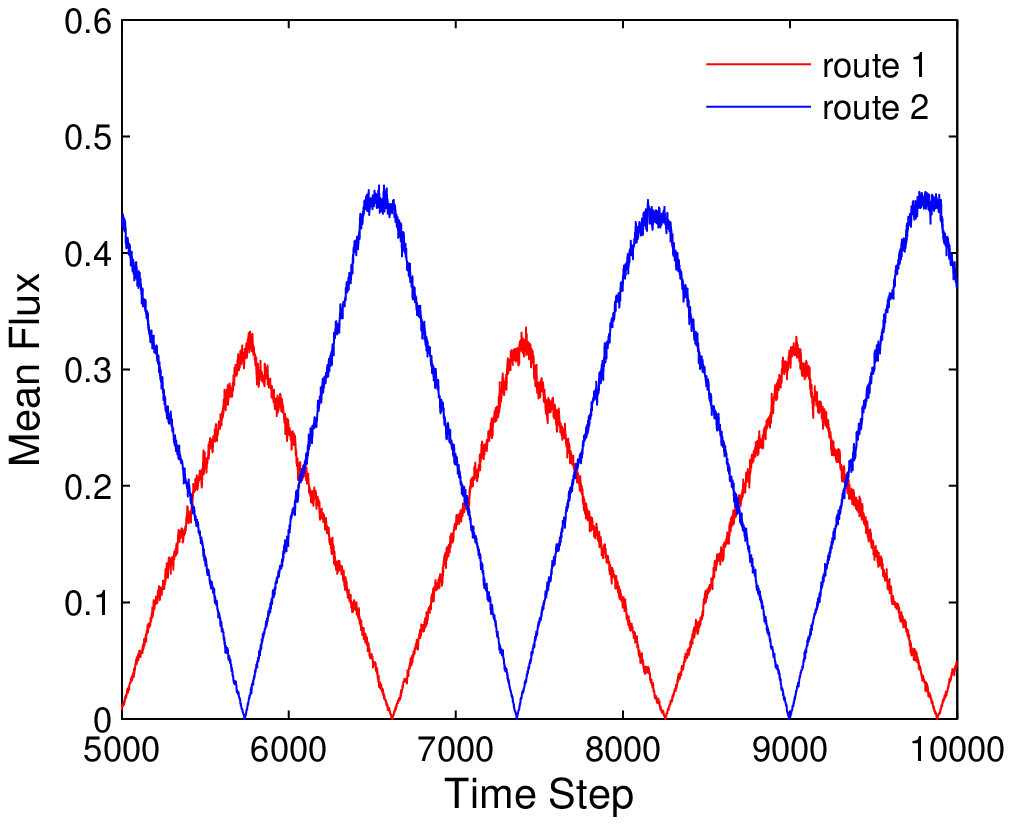}}
    \subfigure[MVFS]{
    \includegraphics[width=2in]{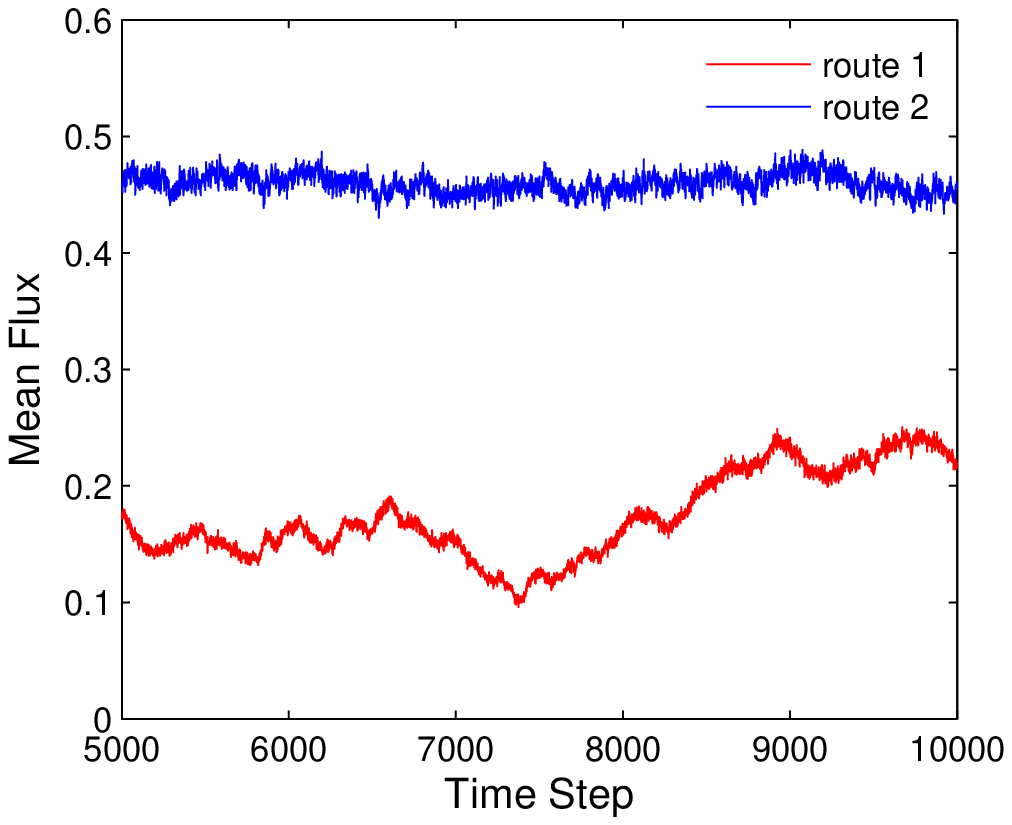}}
    \subfigure[CCFS]{
    \includegraphics[width=2in]{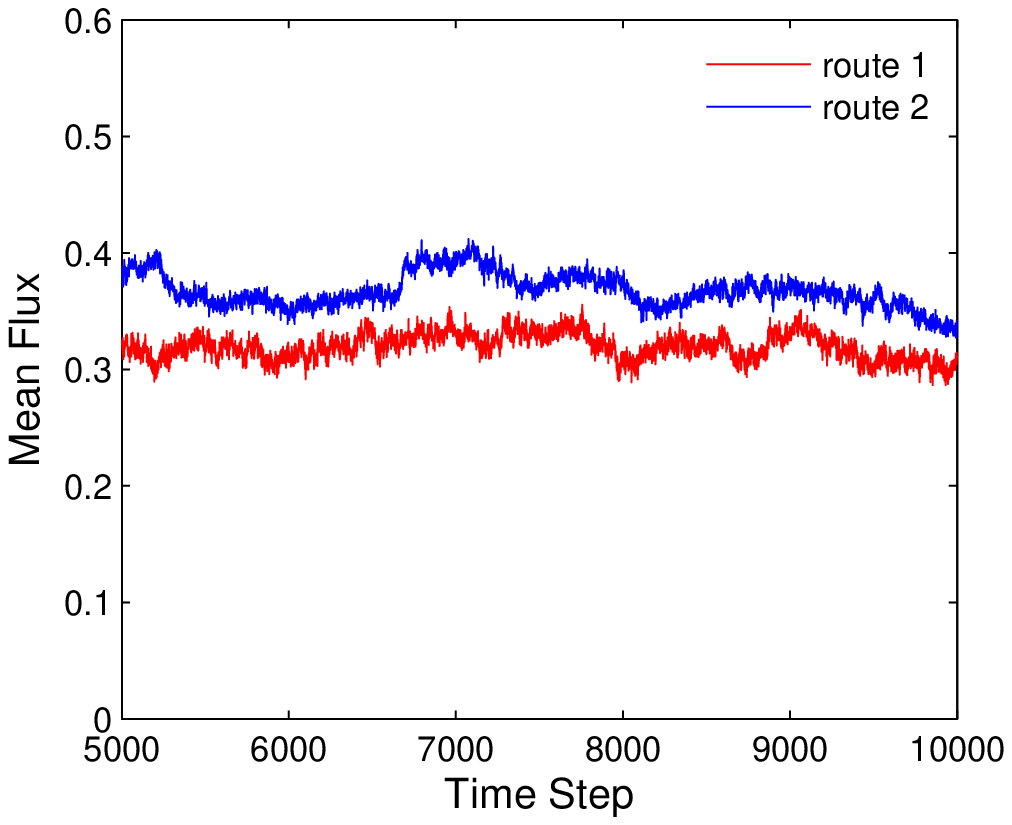}}
    \subfigure[PFS]{
    \includegraphics[width=2in]{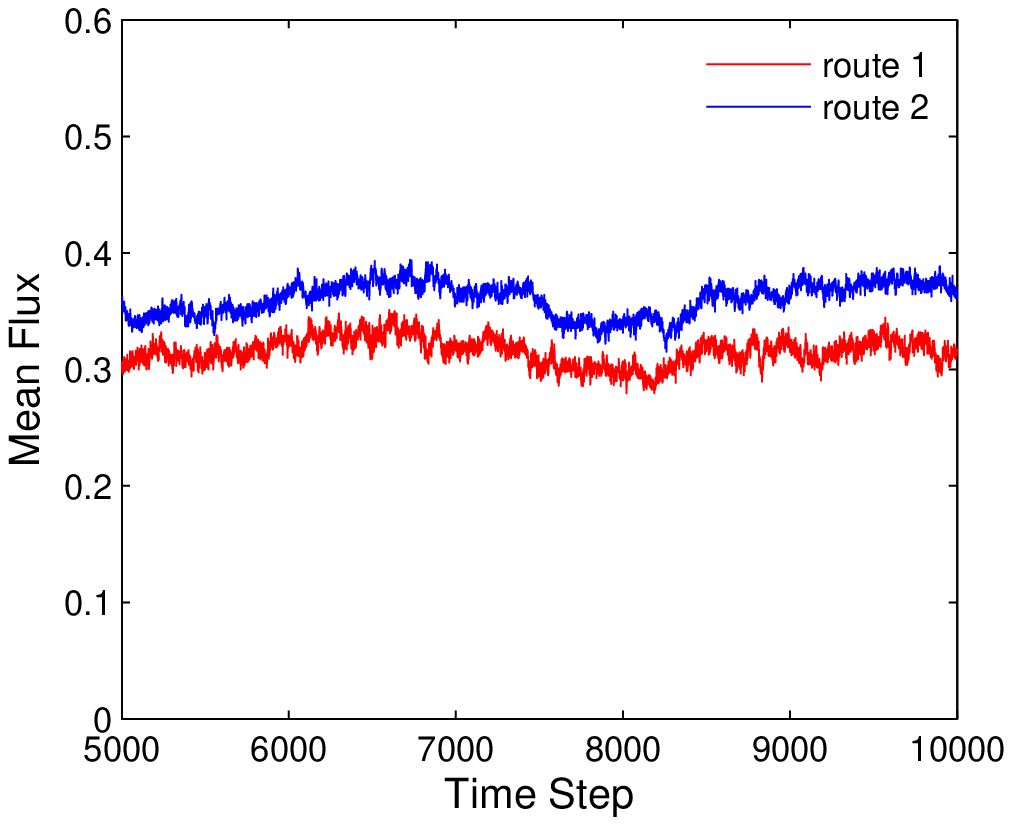}}
    \subfigure[WCCFS]{
    \includegraphics[width=2in]{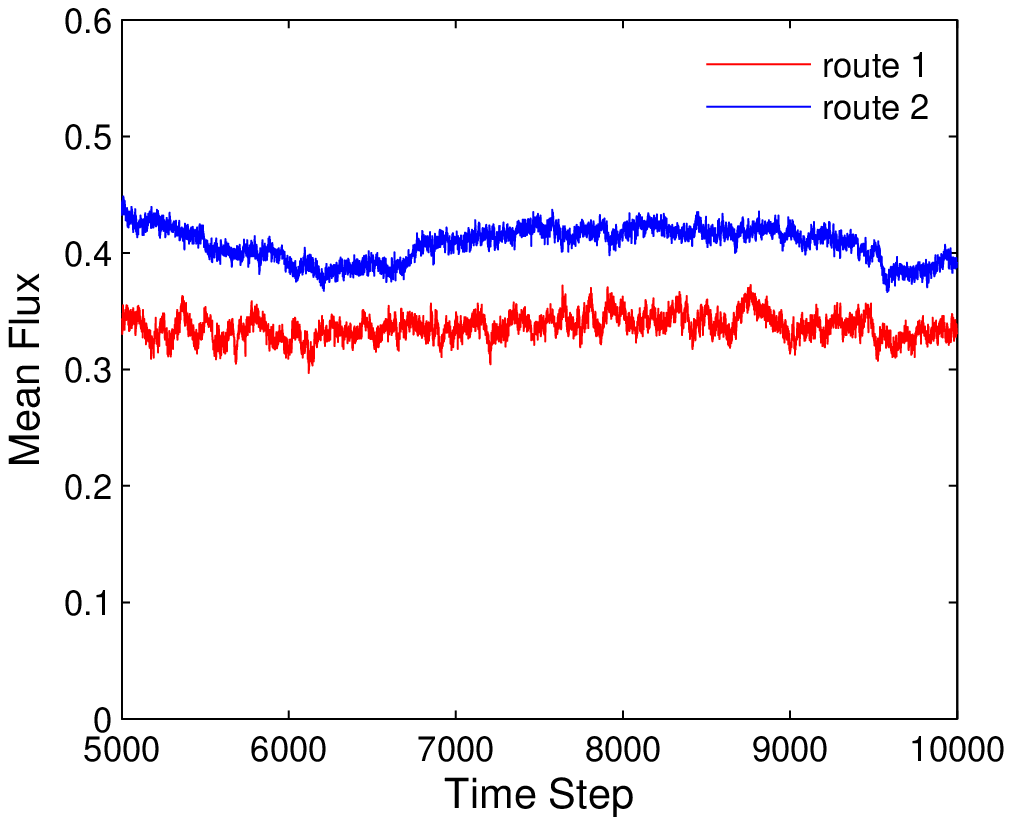}}
    \subfigure[CAFS]{
    \includegraphics[width=2in]{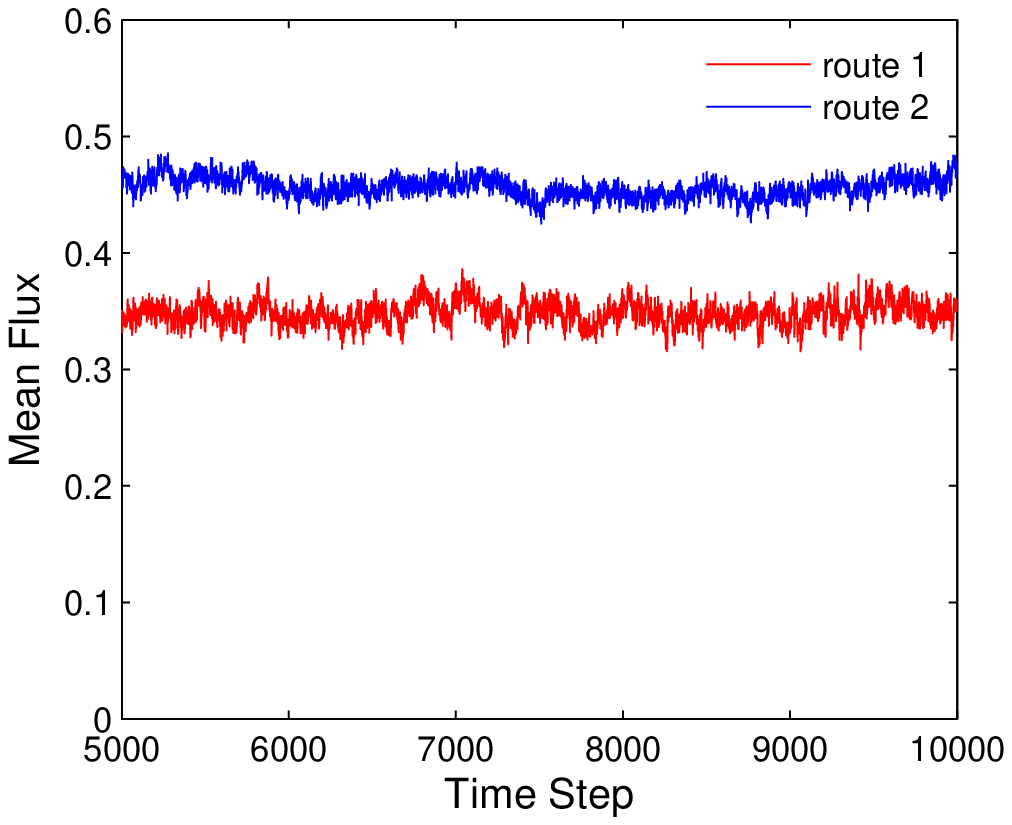}}
    \subfigure[VNFS]{
    \includegraphics[width=2in]{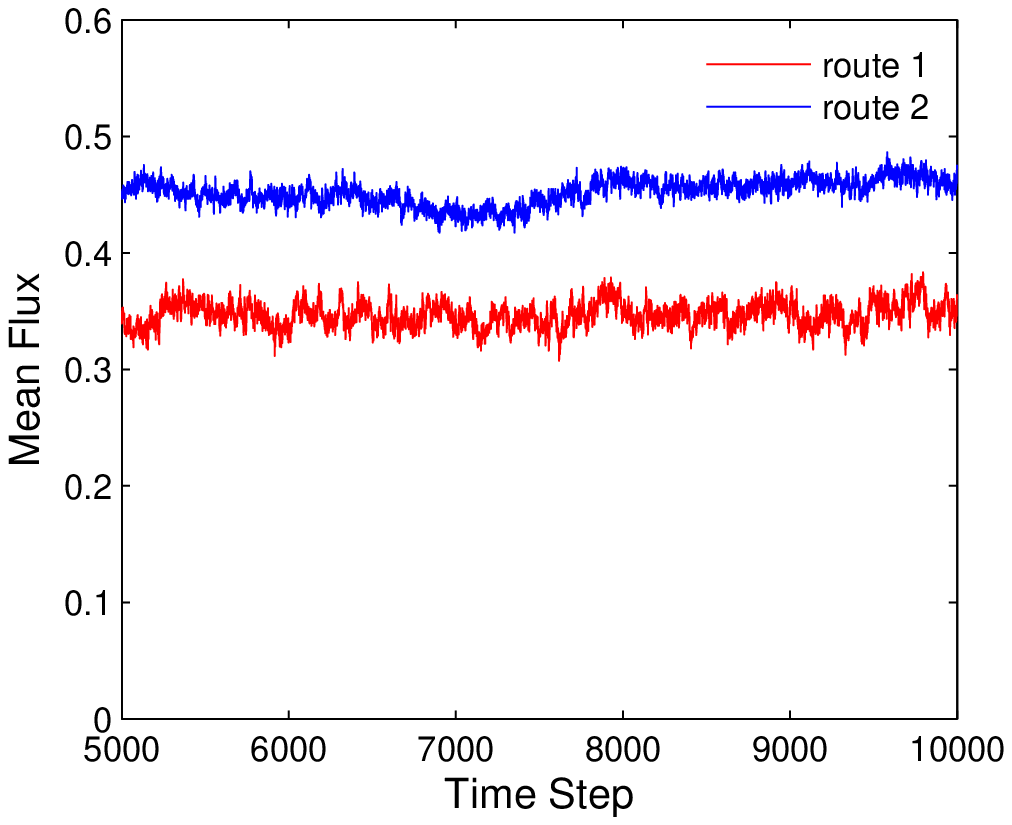}}
    \subfigure[VLFS]{
    \includegraphics[width=2in]{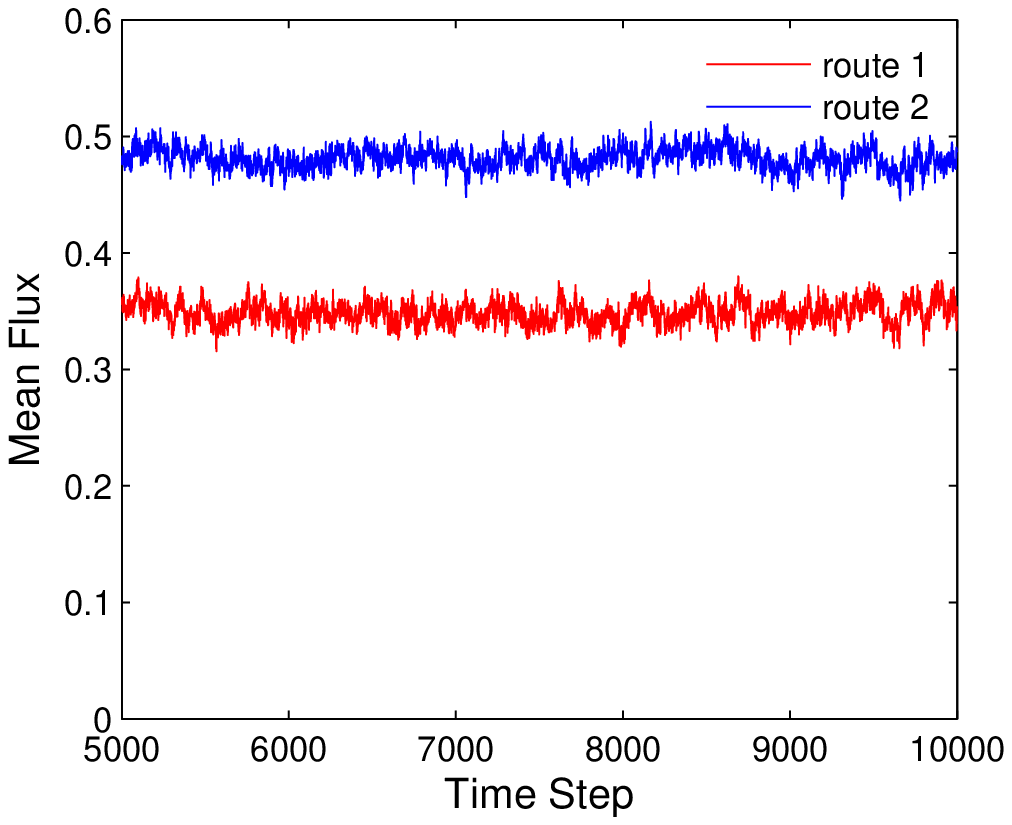}}
    \caption[Mean flux on the asymmetric network.]
    {Mean flux (velocity sum of all vehicles over route length) on the asymmetric network.}
    \label{fig:MeanFlux}
\end{figure}

\begin{figure}
    \centering
    \subfigure[TTFS]{
    \includegraphics[width=2in]{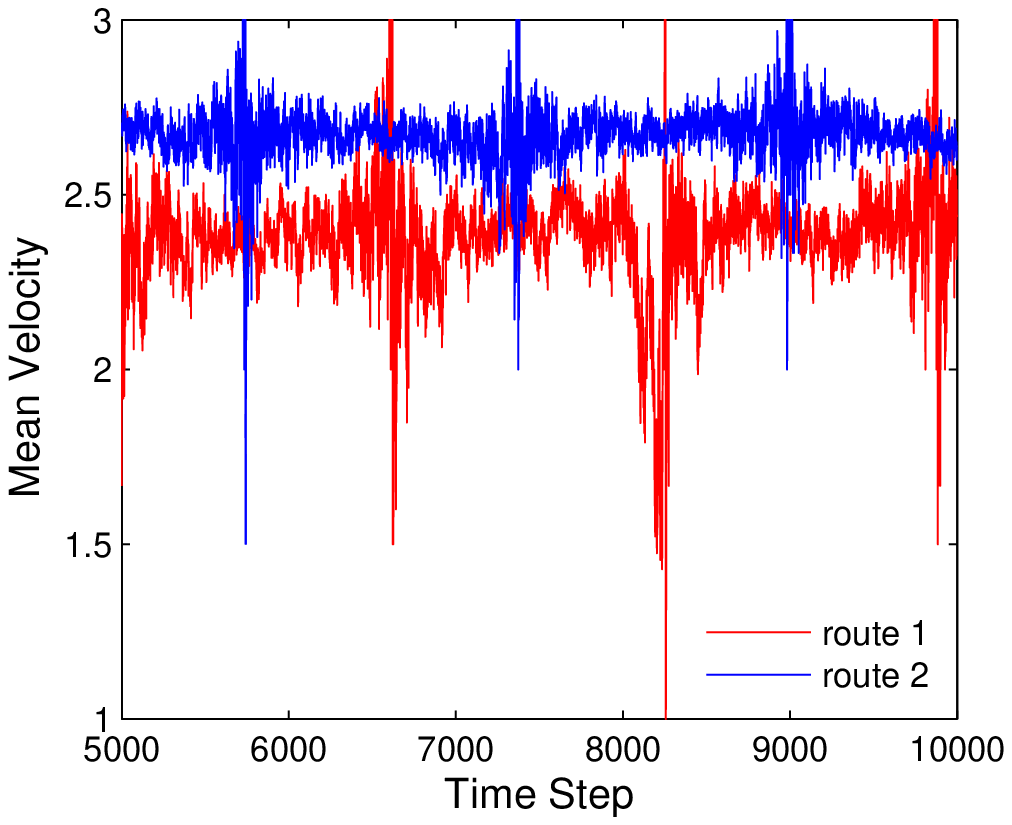}}
    \subfigure[MVFS]{
    \includegraphics[width=2in]{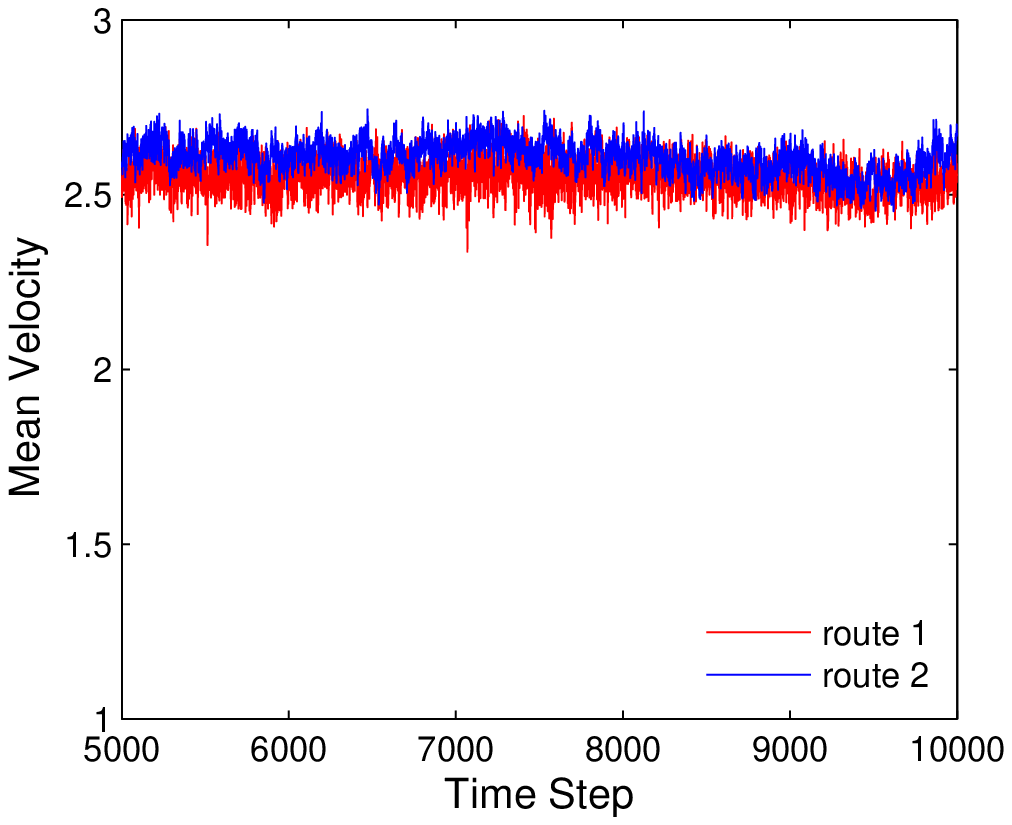}}
    \subfigure[CCFS]{
    \includegraphics[width=2in]{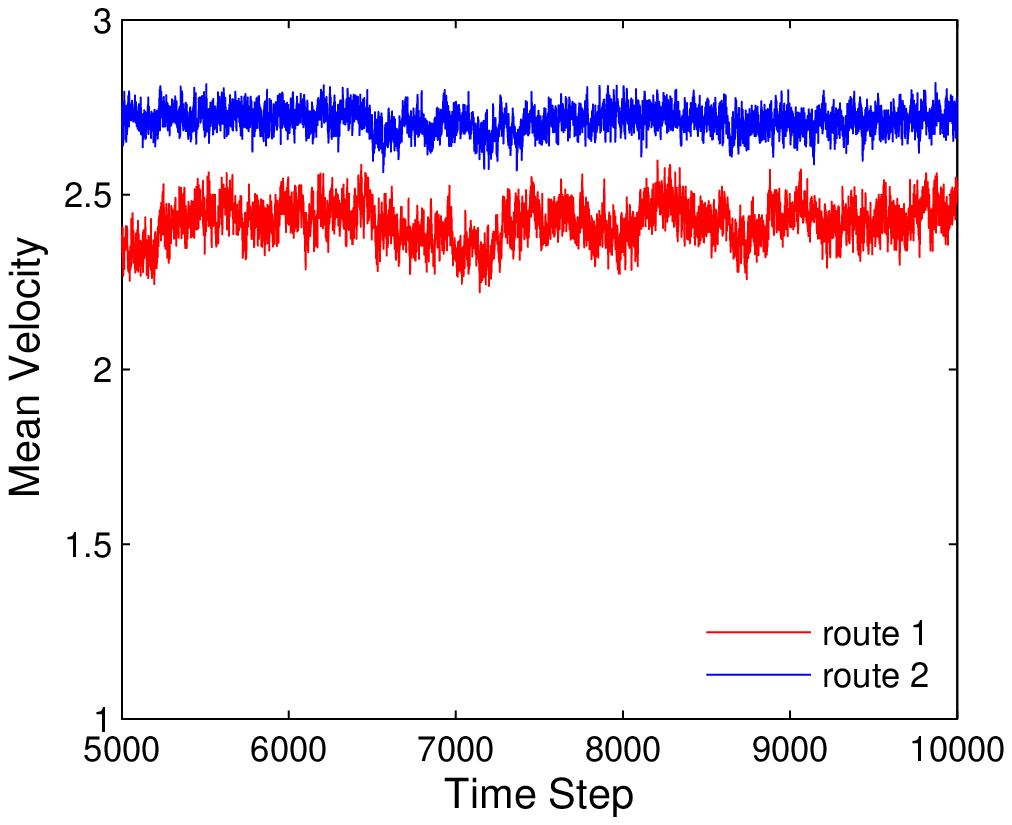}}
    \subfigure[PFS]{
    \includegraphics[width=2in]{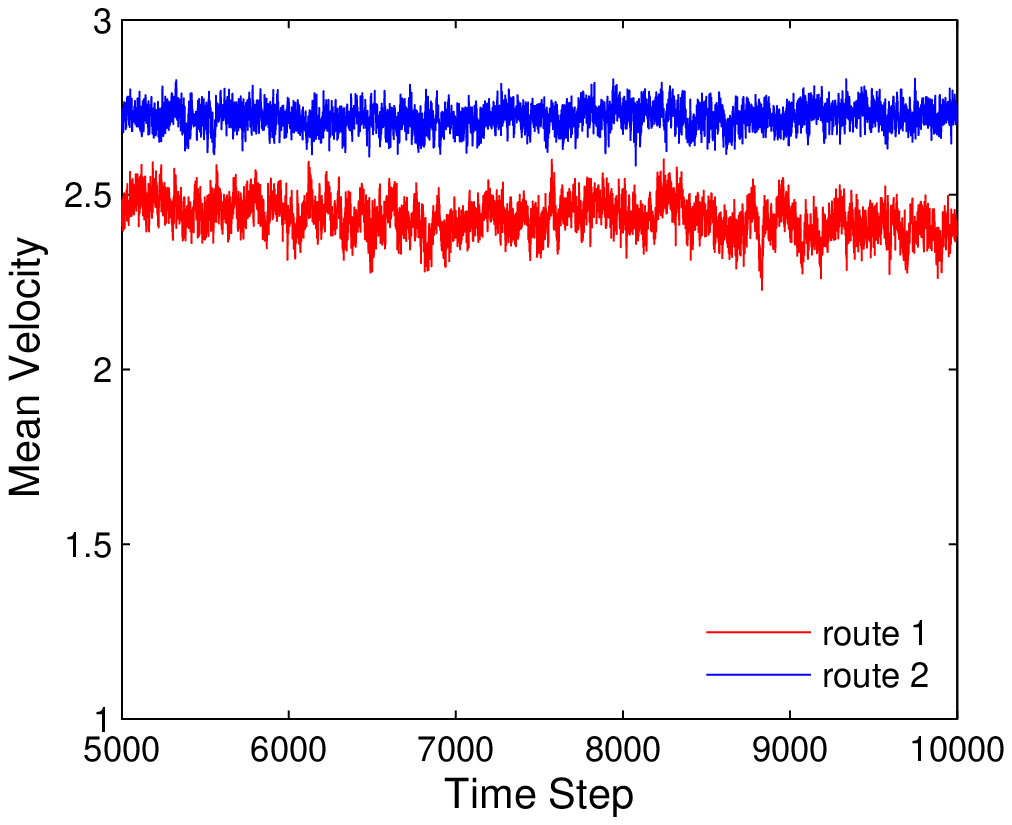}}
    \subfigure[WCCFS]{
    \includegraphics[width=2in]{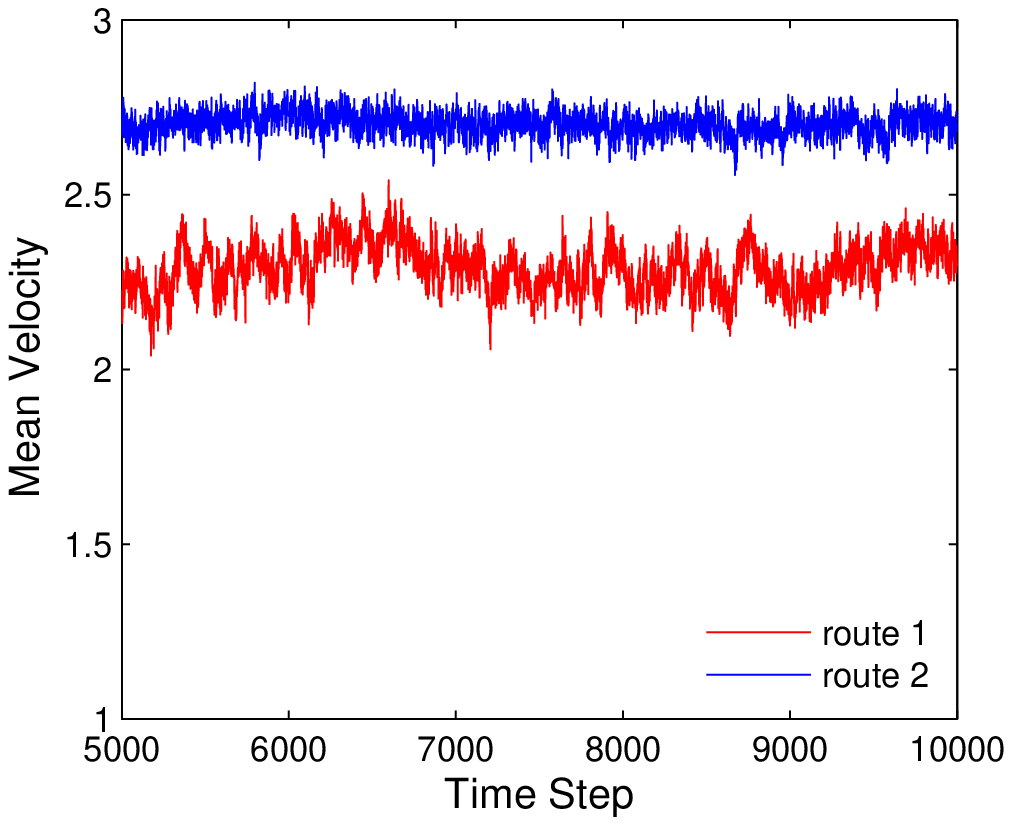}}
    \subfigure[CAFS]{
    \includegraphics[width=2in]{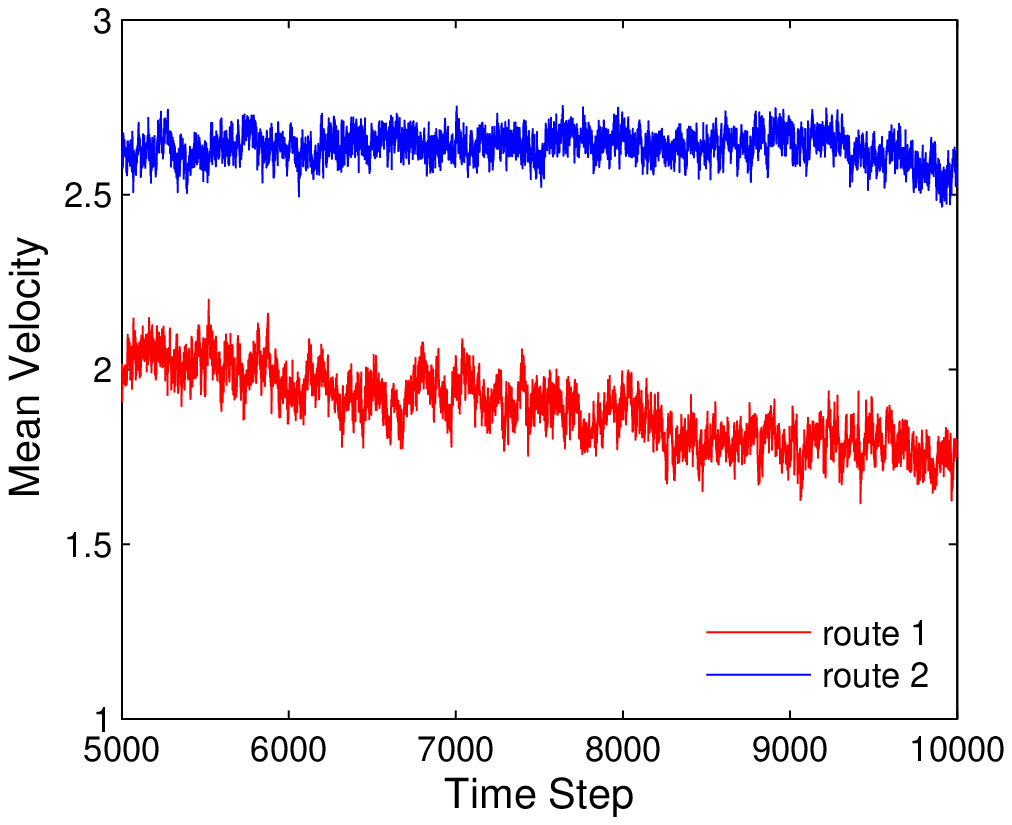}}
    \subfigure[VNFS]{
    \includegraphics[width=2in]{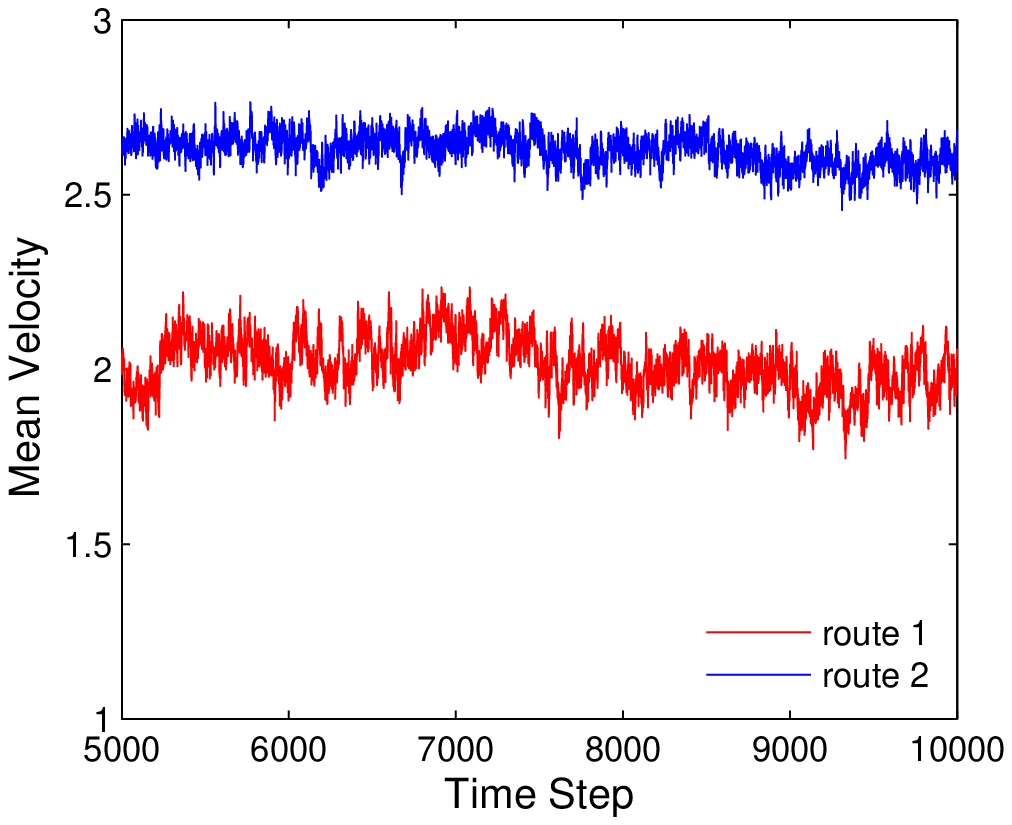}}
    \subfigure[VLFS]{
    \includegraphics[width=2in]{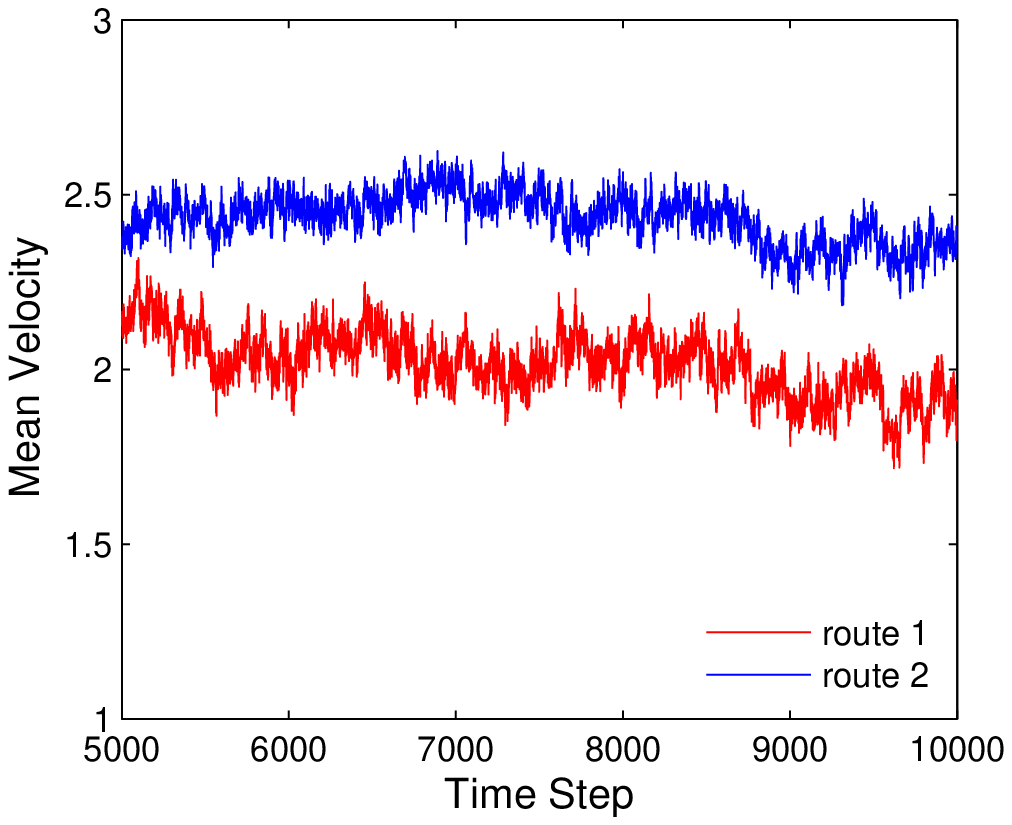}}
    \caption{Mean velocity on the asymmetric network.}
    \label{fig:MeanVelocity}
\end{figure}

\begin{figure}
    \centering
    \subfigure[TTFS]{
    \includegraphics[width=2in]{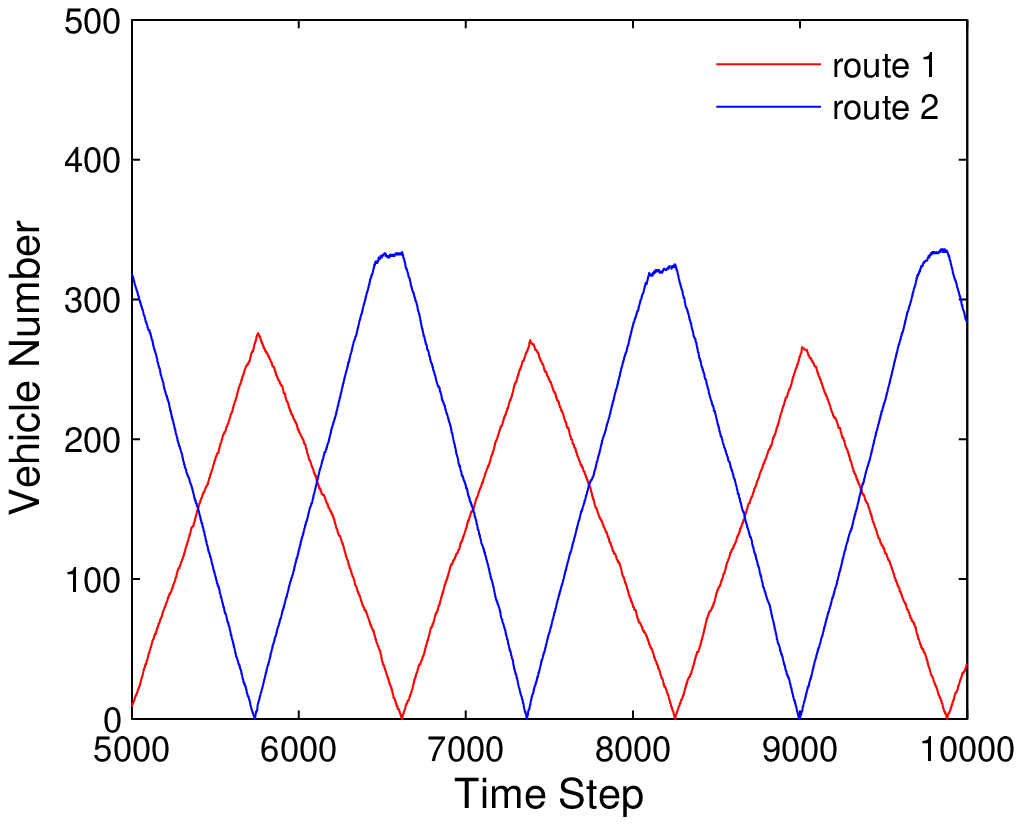}}
    \subfigure[MVFS]{
    \includegraphics[width=2in]{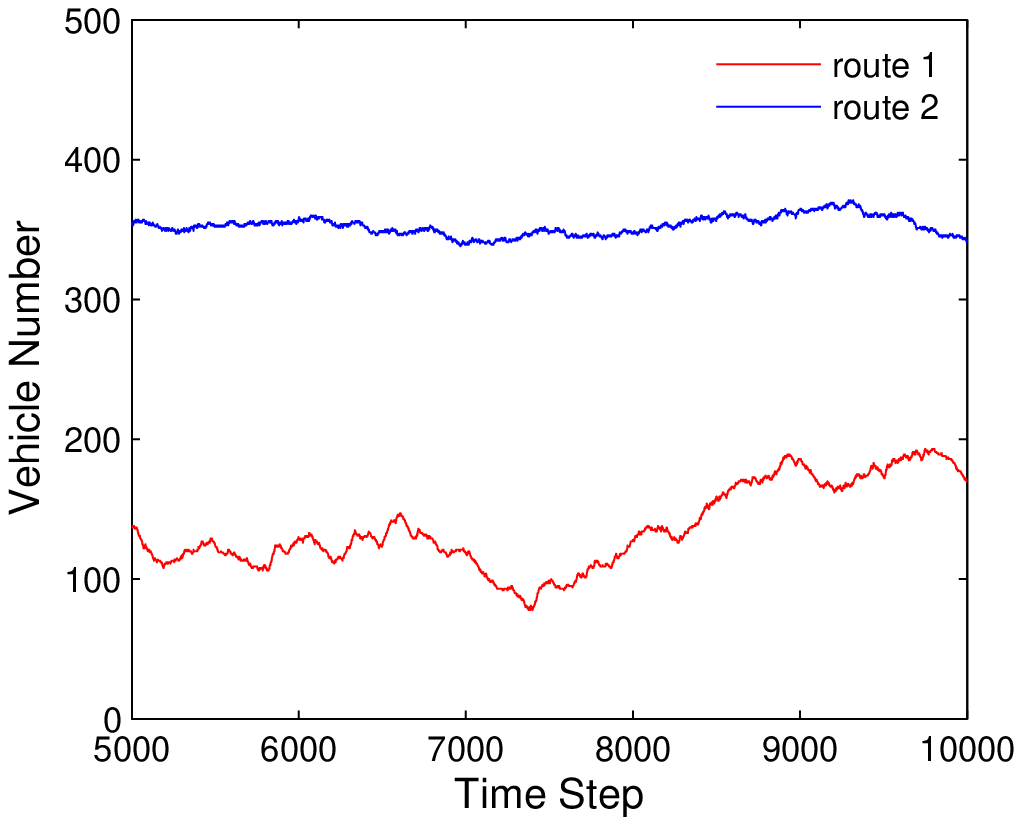}}
    \subfigure[CCFS]{
    \includegraphics[width=2in]{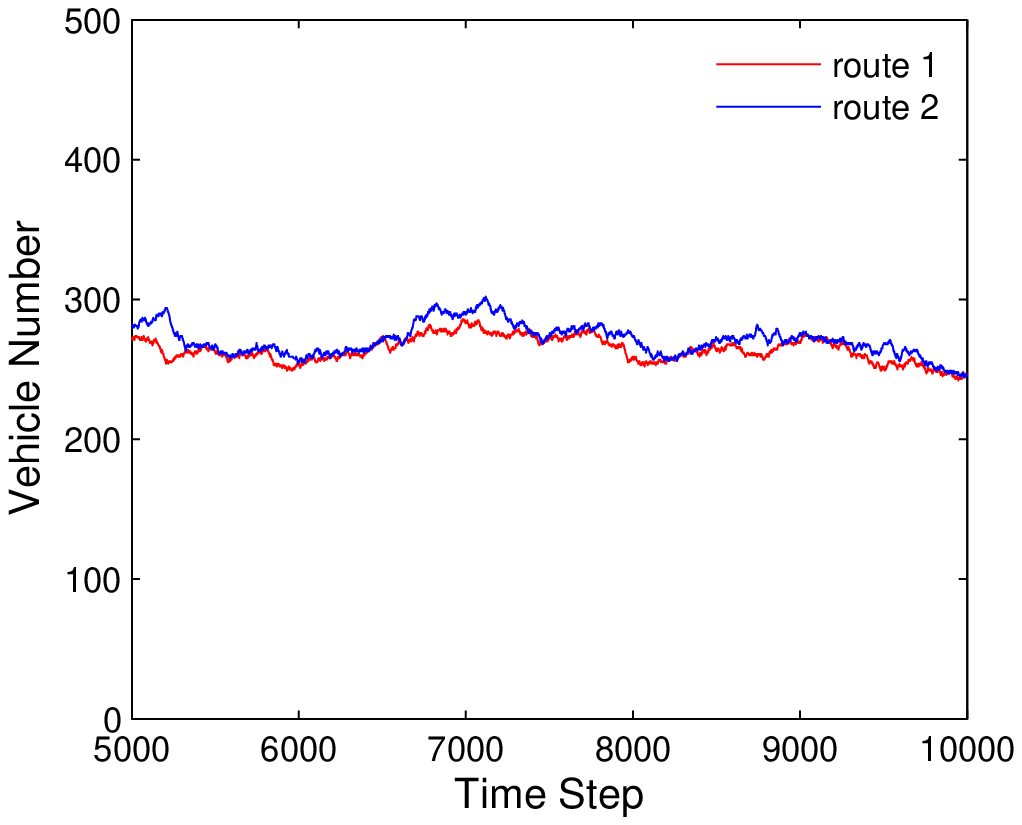}}
    \subfigure[PFS]{
    \includegraphics[width=2in]{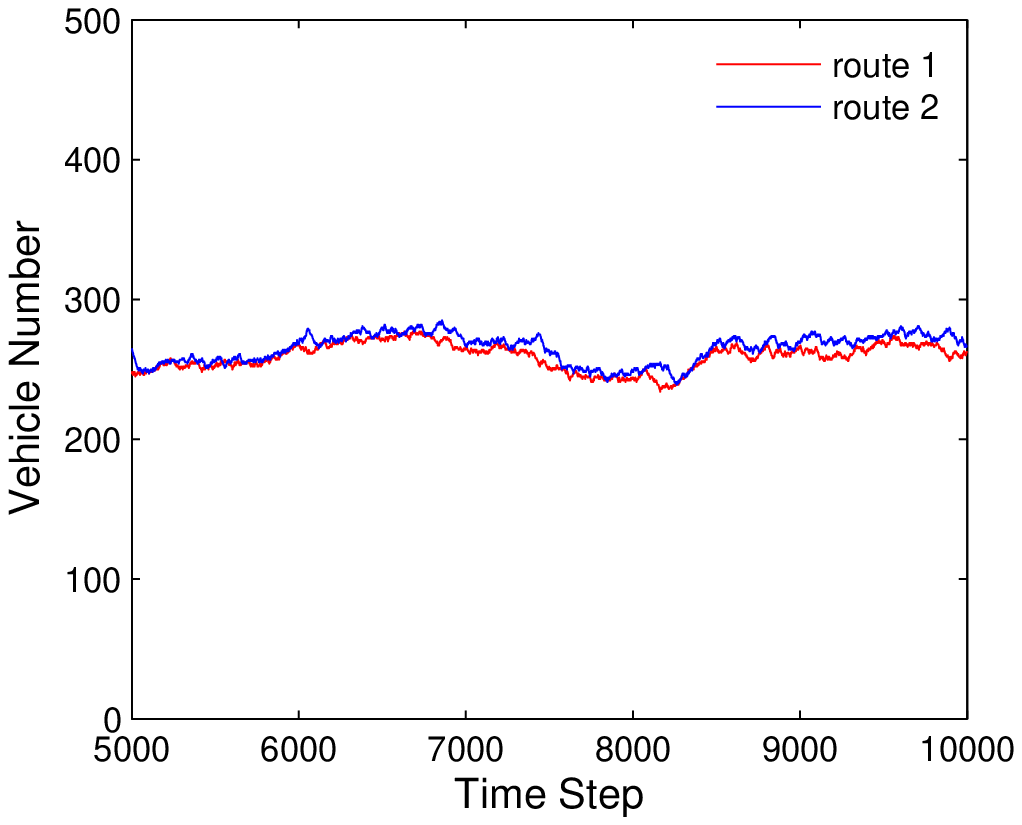}}
    \subfigure[WCCFS]{
    \includegraphics[width=2in]{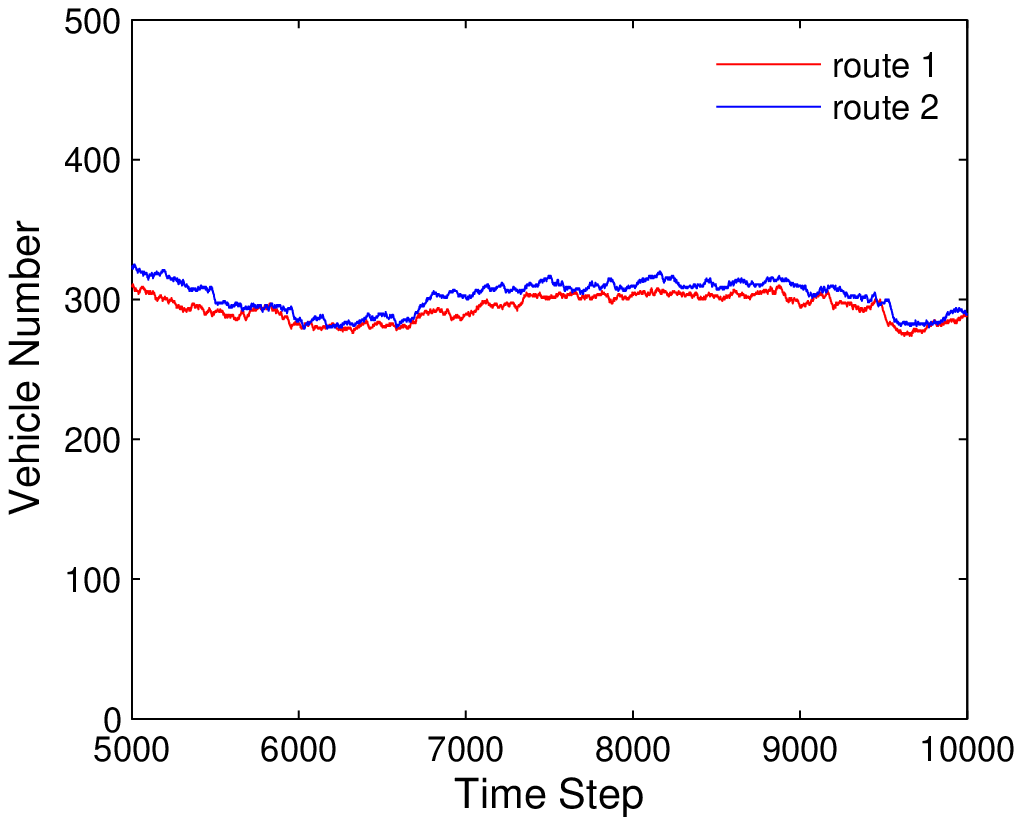}}
    \subfigure[CAFS]{
    \includegraphics[width=2in]{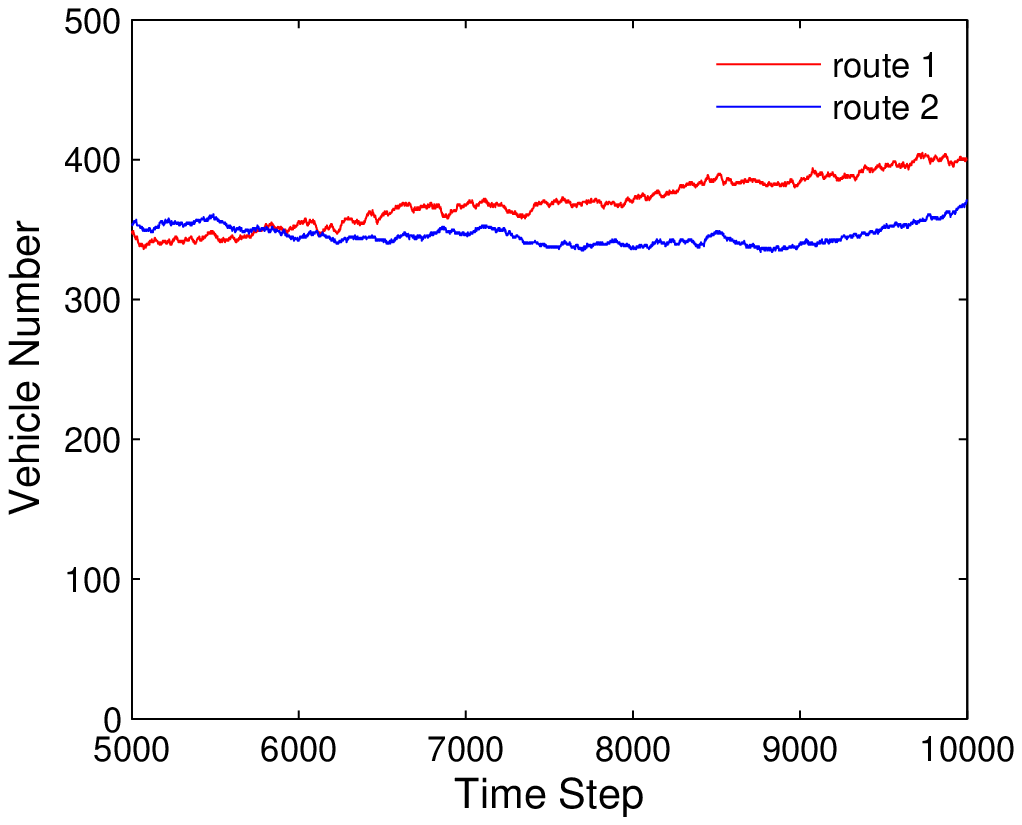}}
    \subfigure[VNFS]{
    \includegraphics[width=2in]{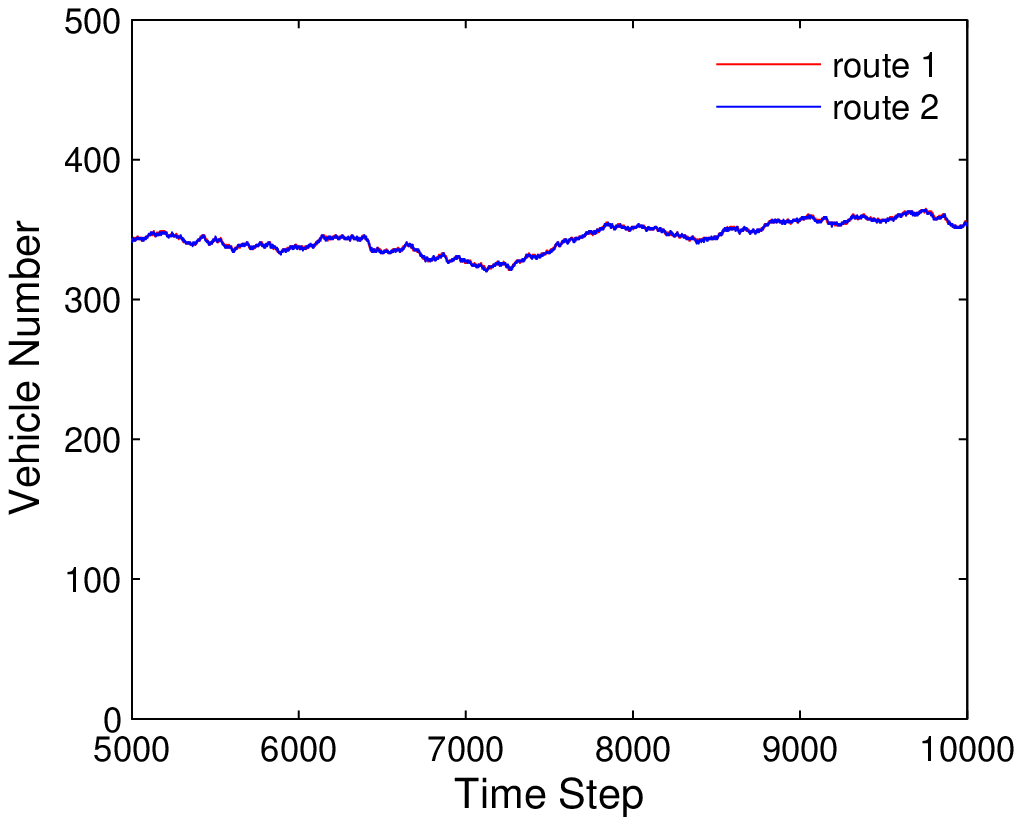}}
    \subfigure[VLFS]{
    \includegraphics[width=2in]{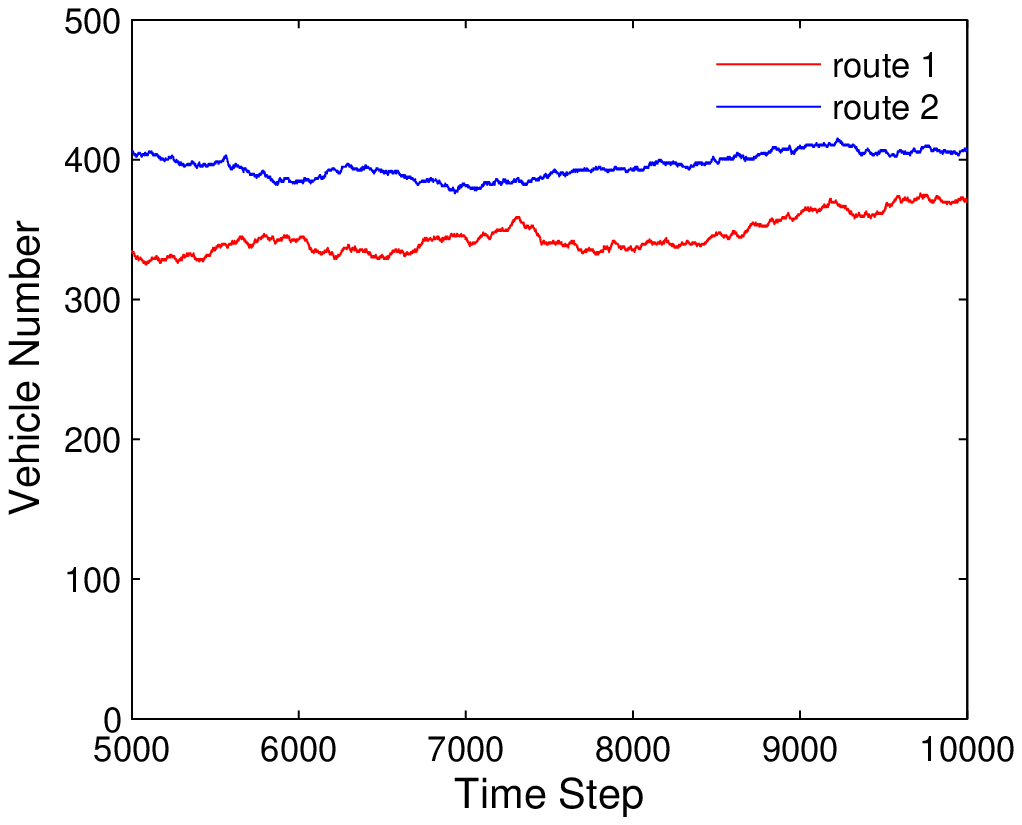}}
    \caption{Vehicle numbers on the asymmetric network diverting.}
    \label{fig:VehicleNumber}
\end{figure}

Figure \ref{fig:TravelTime} compares the travel times resulted by the eight route guidance strategies on two routes at all time steps.
Figure \ref{fig:MeanFlux}, \ref{fig:MeanVelocity} and \ref{fig:VehicleNumber} present the corresponding traffic conditions.
It can be seen that only MVFS results in close travel times on two routes, i.e., approximates UO, while the others fail.
The reason is straightforward that only mean velocity in MVFS has a close relation with travel time, i.e., close mean velocity leads to close travel time on alternative routes; the congestion coefficient, vehicle number, vacancy length, etc. in the other strategies are unable to reflect the relations in the considered scenario.
Thus, only MVFS approximates UO by equalizing mean velocity.

\begin{figure}
    \centering
    \subfigure[TTFS]{
    \includegraphics[width=2in]{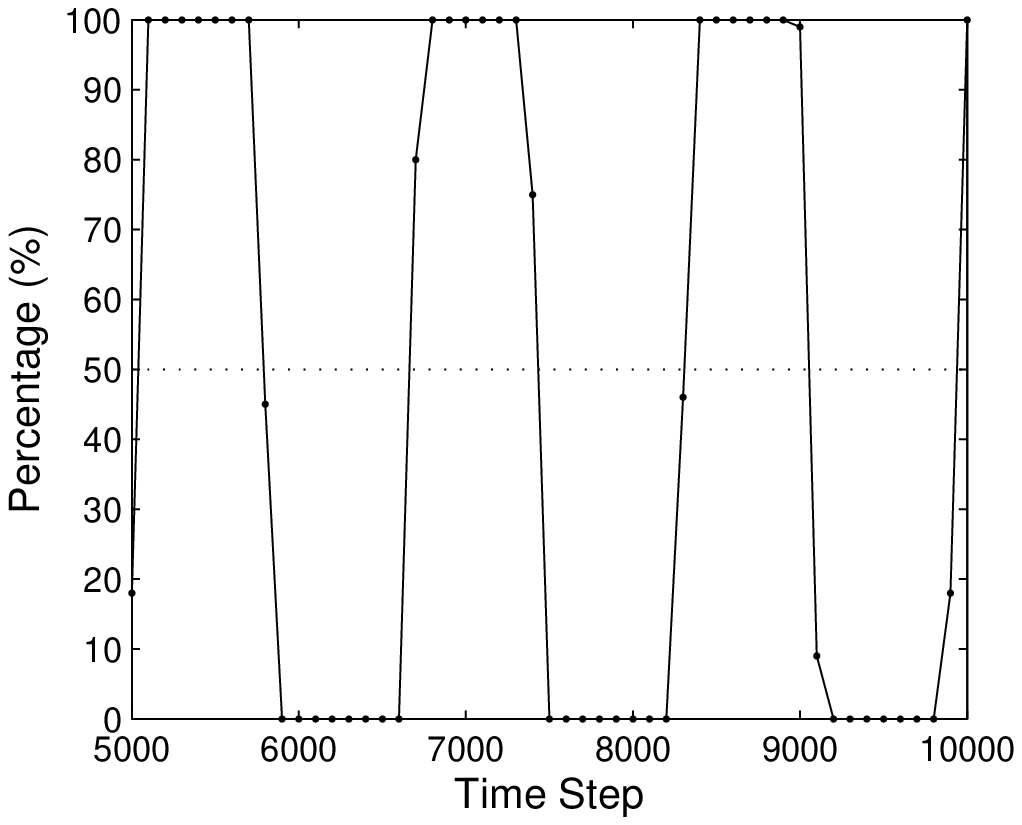}}
    \subfigure[MVFS]{
    \includegraphics[width=2in]{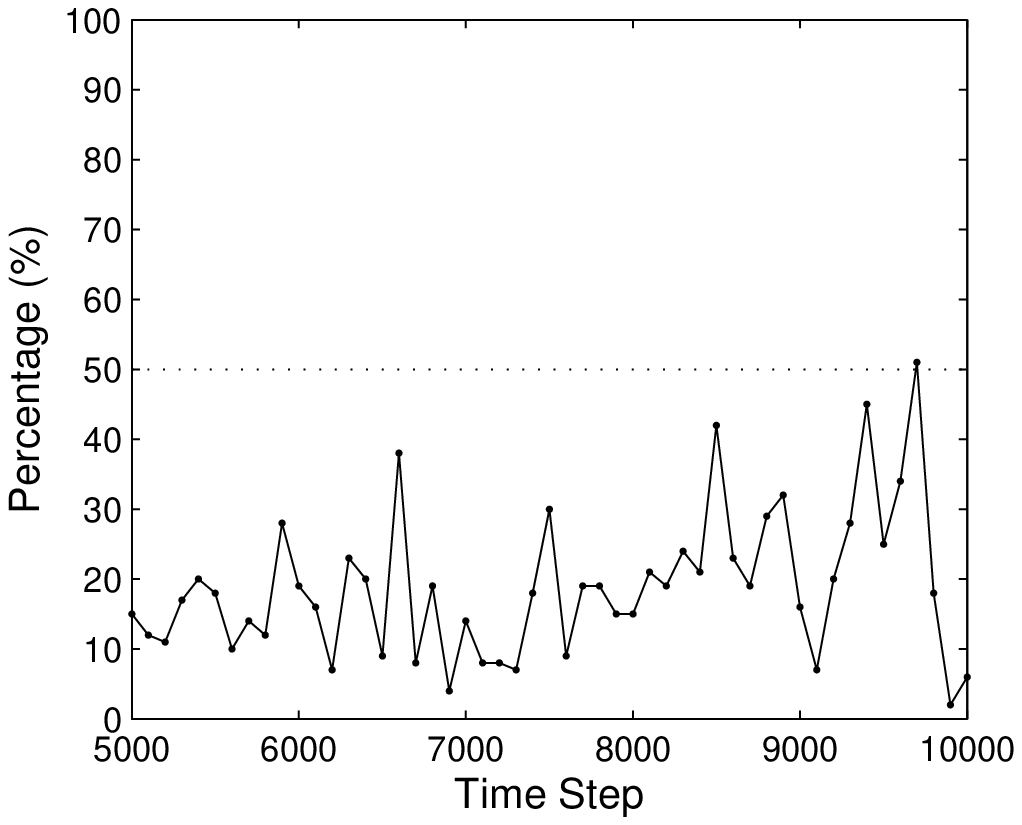}}
    \subfigure[CCFS]{
    \includegraphics[width=2in]{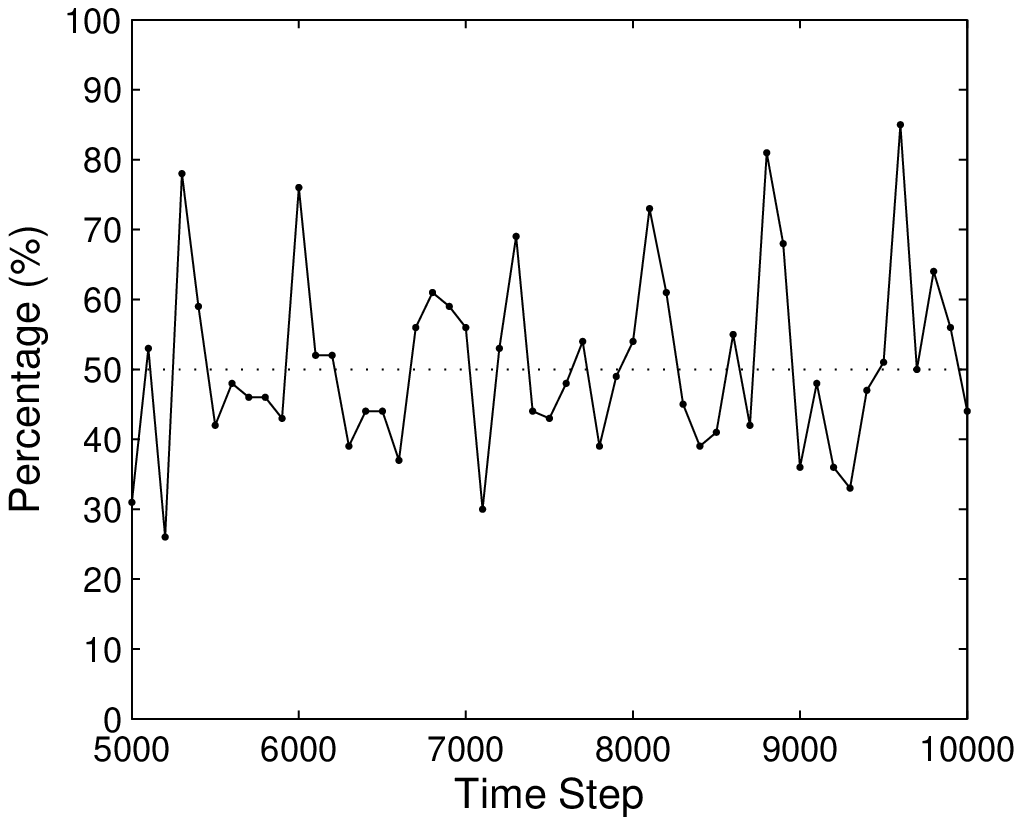}}
    \subfigure[PFS]{
    \includegraphics[width=2in]{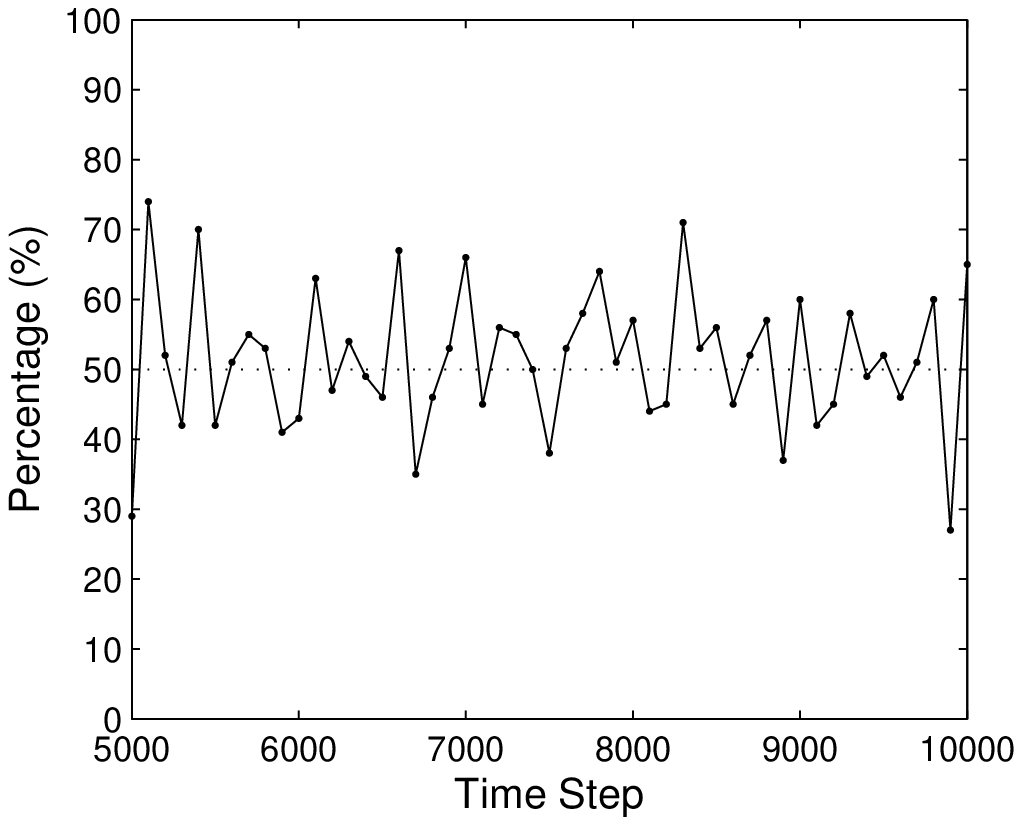}}
    \subfigure[WCCFS]{
    \includegraphics[width=2in]{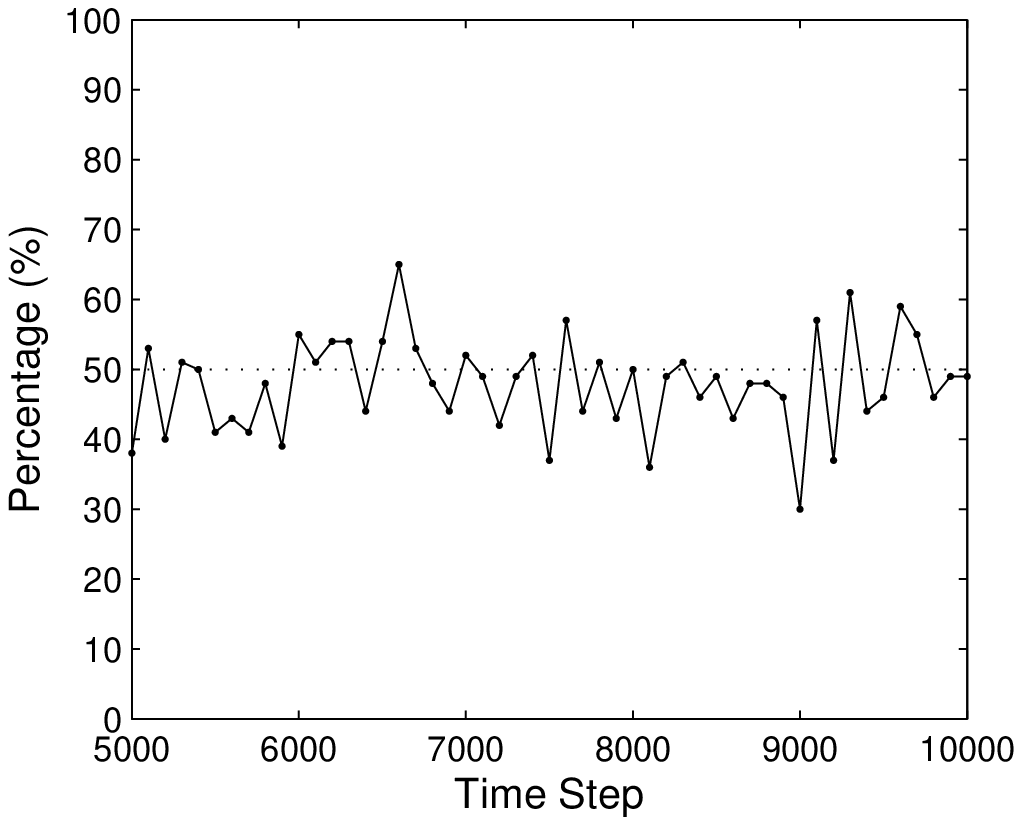}}
    \subfigure[CAFS]{
    \includegraphics[width=2in]{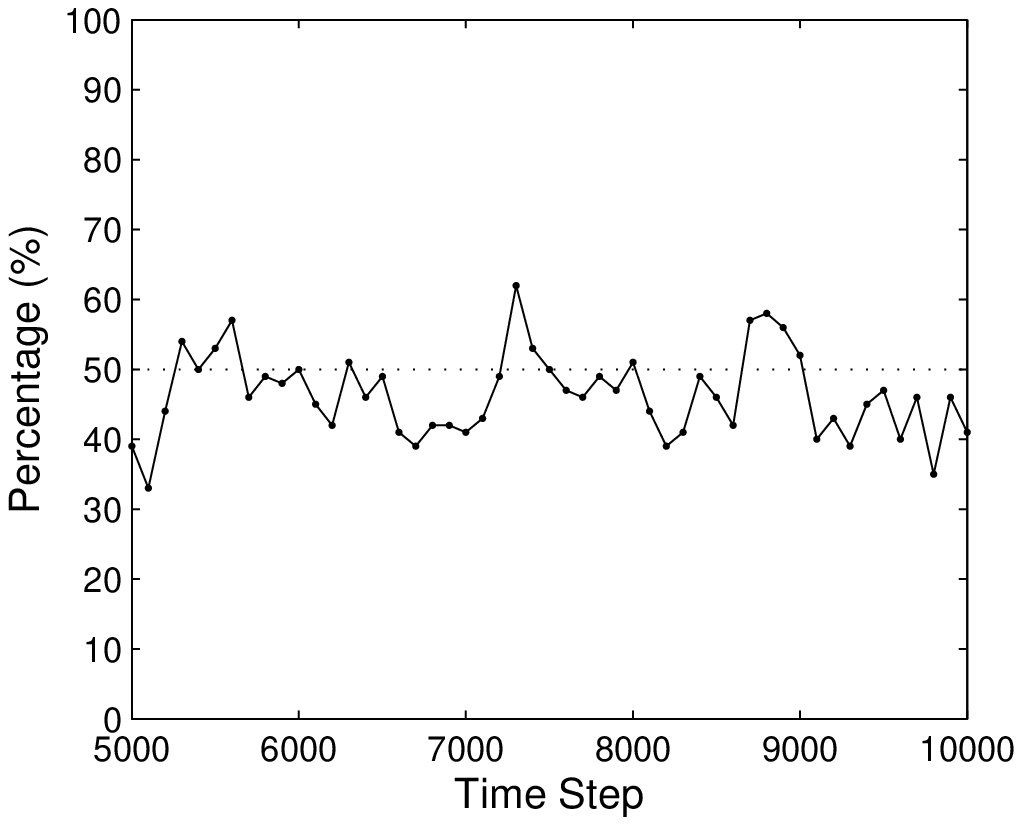}}
    \subfigure[VNFS]{
    \includegraphics[width=2in]{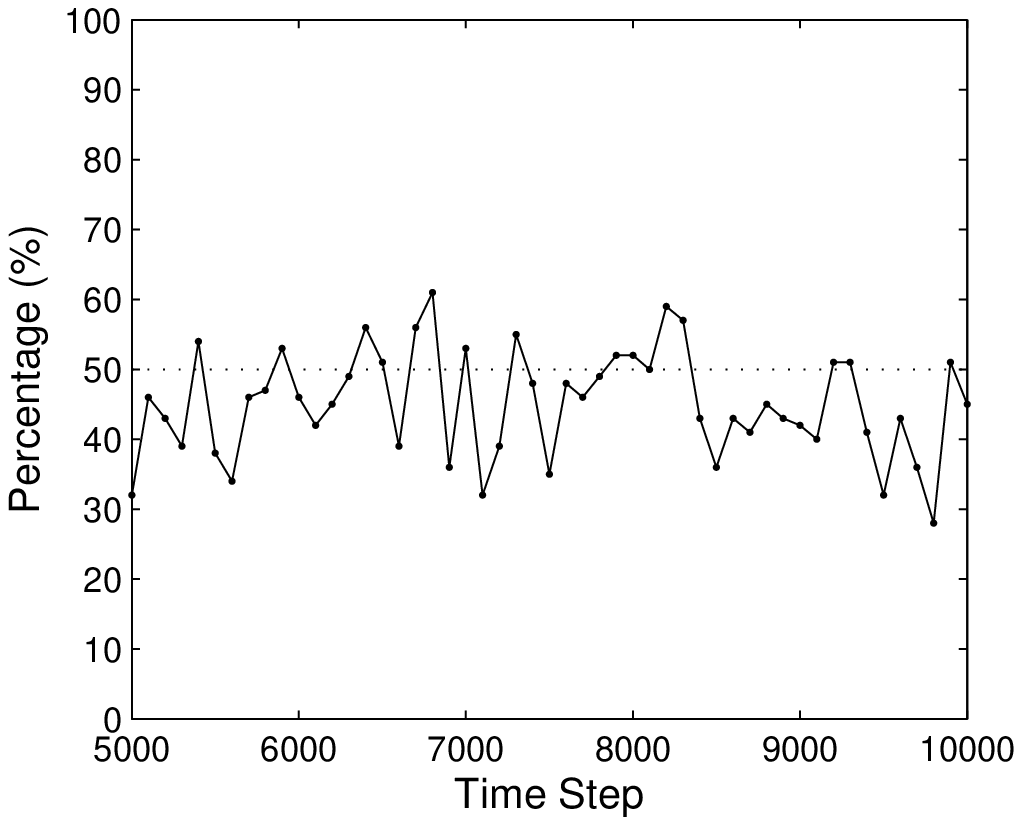}}
    \subfigure[VLFS]{
    \includegraphics[width=2in]{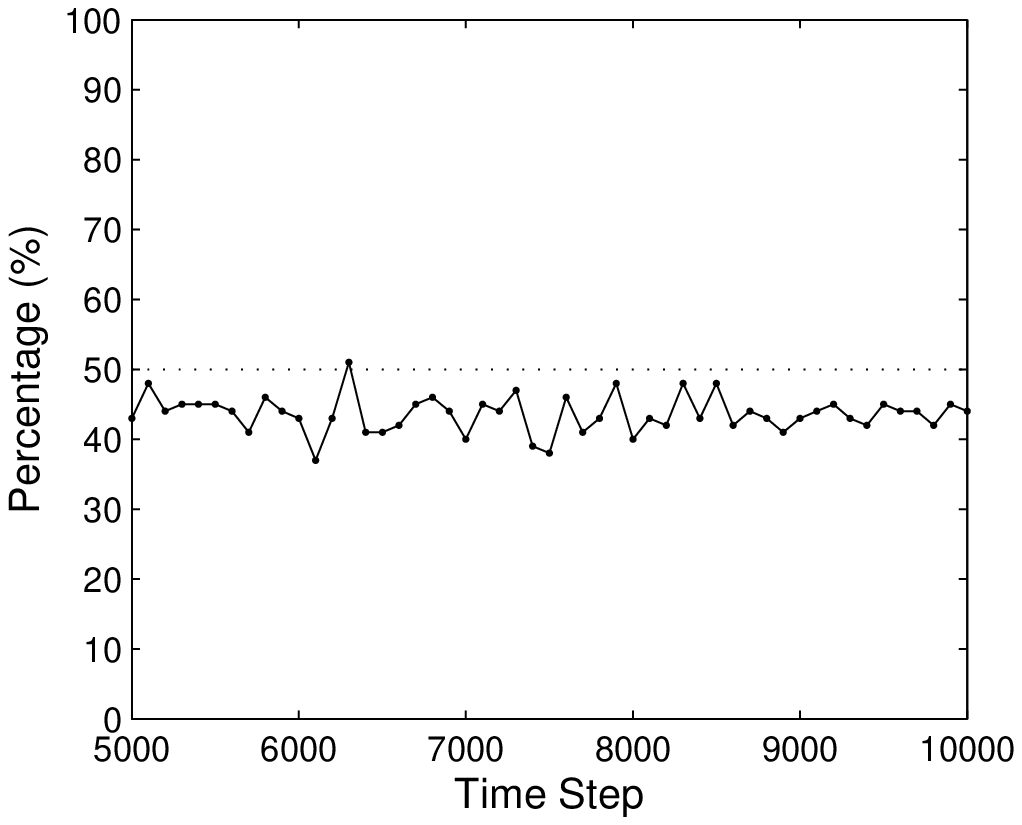}}
    \caption[Vehicle percentages diverting on route 1 on the asymmetric network]
            {Vehicle percentages diverting on route 1 on the asymmetric network
            (each rate is measured within 100 time steps).}
    \label{fig:SplitRate}
\end{figure}

To show the ability of reflecting the difference of slowdown probability,
Figure \ref{fig:SplitRate} presents the diversion rates of the eight strategies.
Except for TTFS and MVFS, all strategies still divert incoming vehicles by around 50\%,
which are similar with those in the symmetric network;
it implies that the strategies are unable to reflect the difference of the slowdown probability.
In contrast, MVFS diverts more vehicles on route 2 due to the worse traffic conditions on route 1.

\begin{figure}
    \centering
    \subfigure[TTFS]{
    \includegraphics[width=2in]{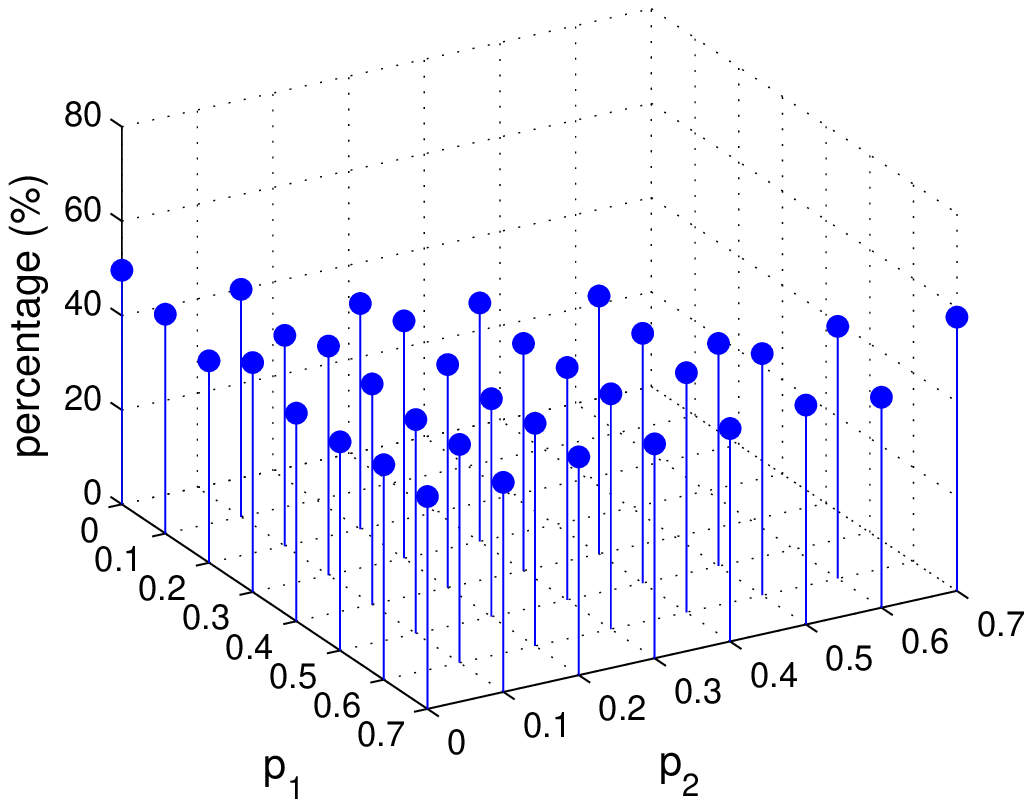}}
    \subfigure[MVFS]{
    \includegraphics[width=2in]{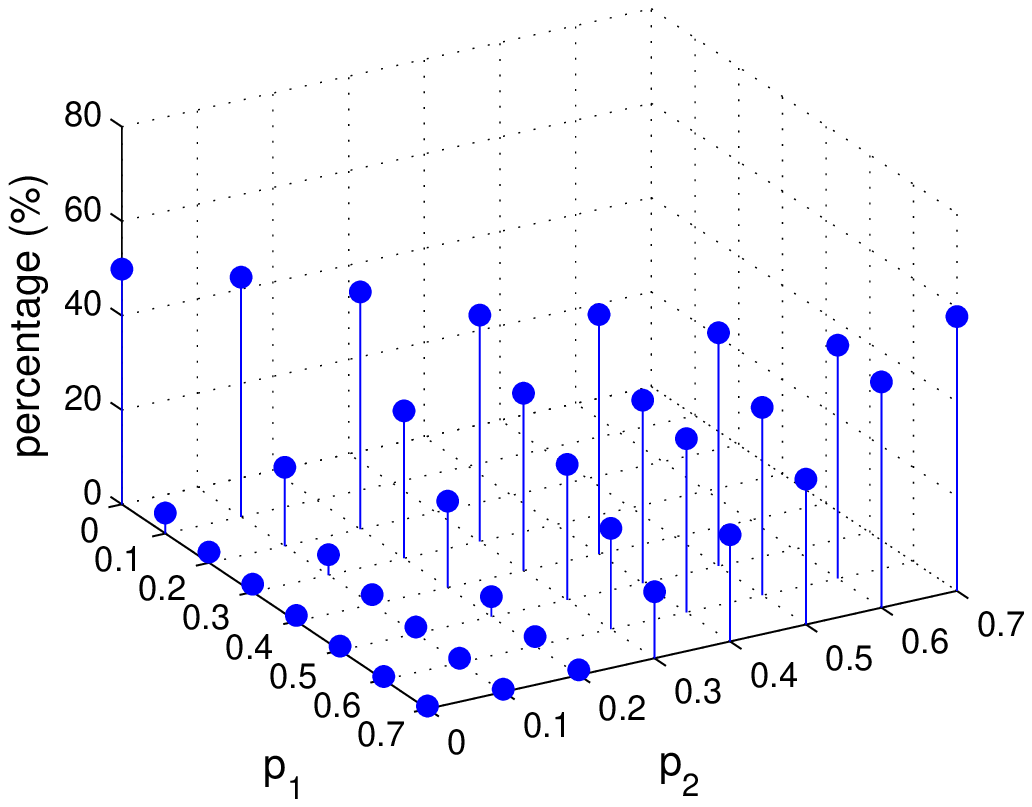}}
    \subfigure[CCFS]{
    \includegraphics[width=2in]{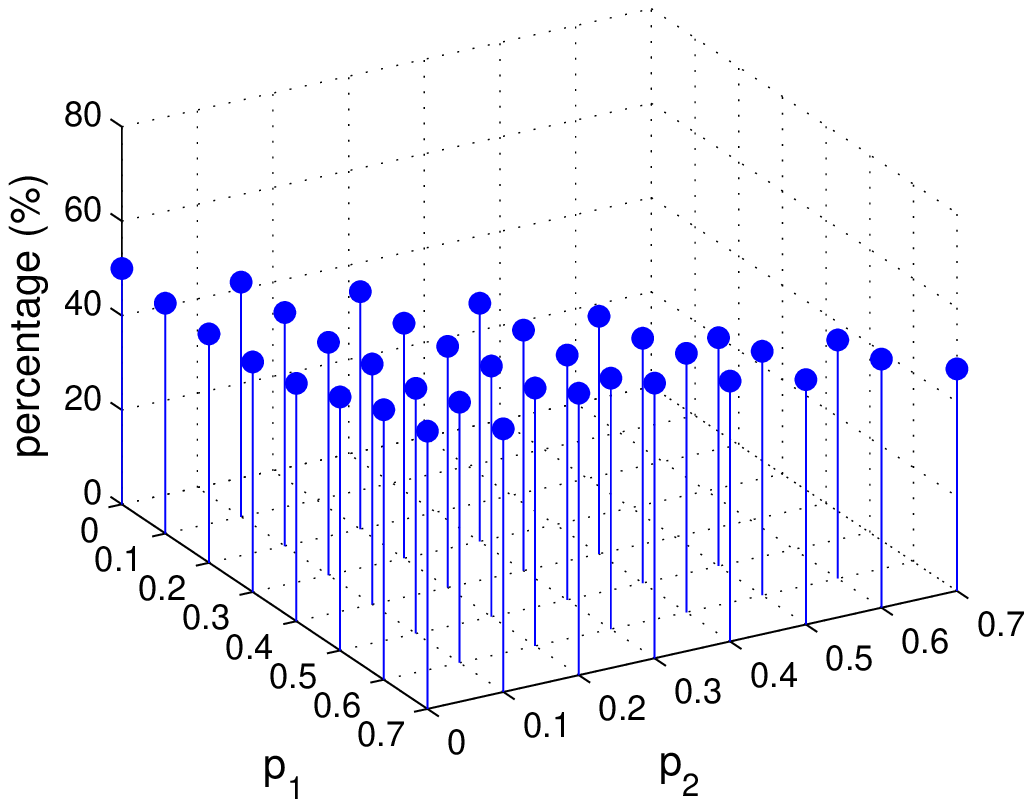}}
    \subfigure[PFS]{
    \includegraphics[width=2in]{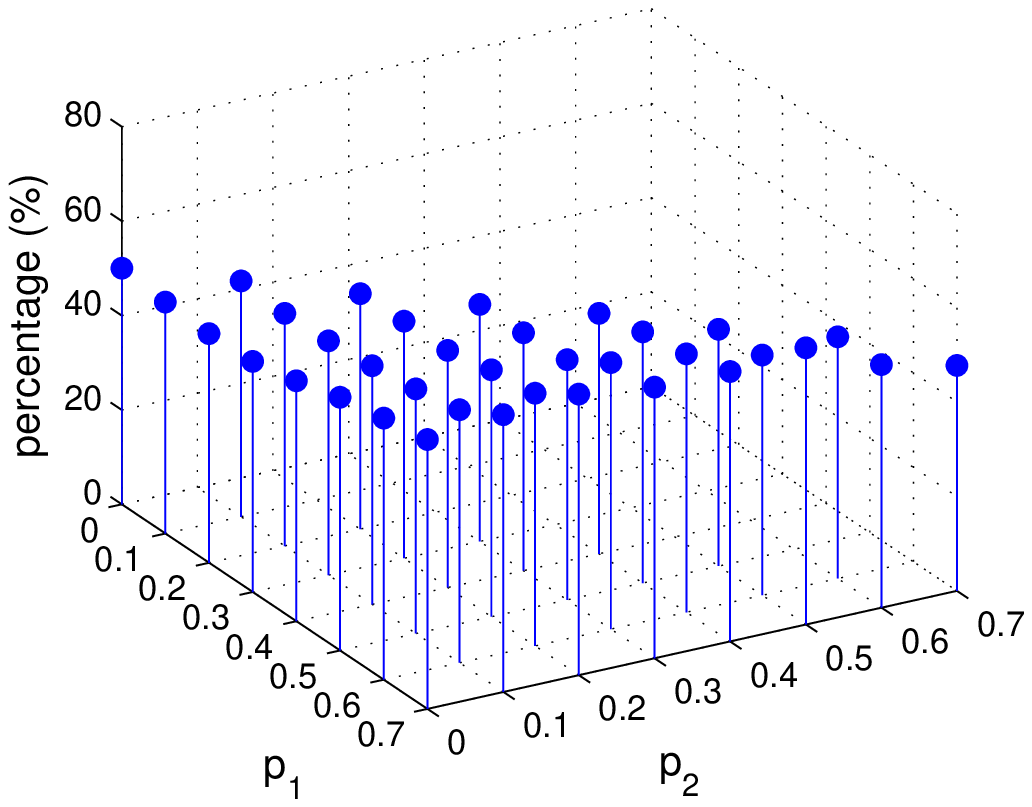}}
    \subfigure[WCCFS]{
    \includegraphics[width=2in]{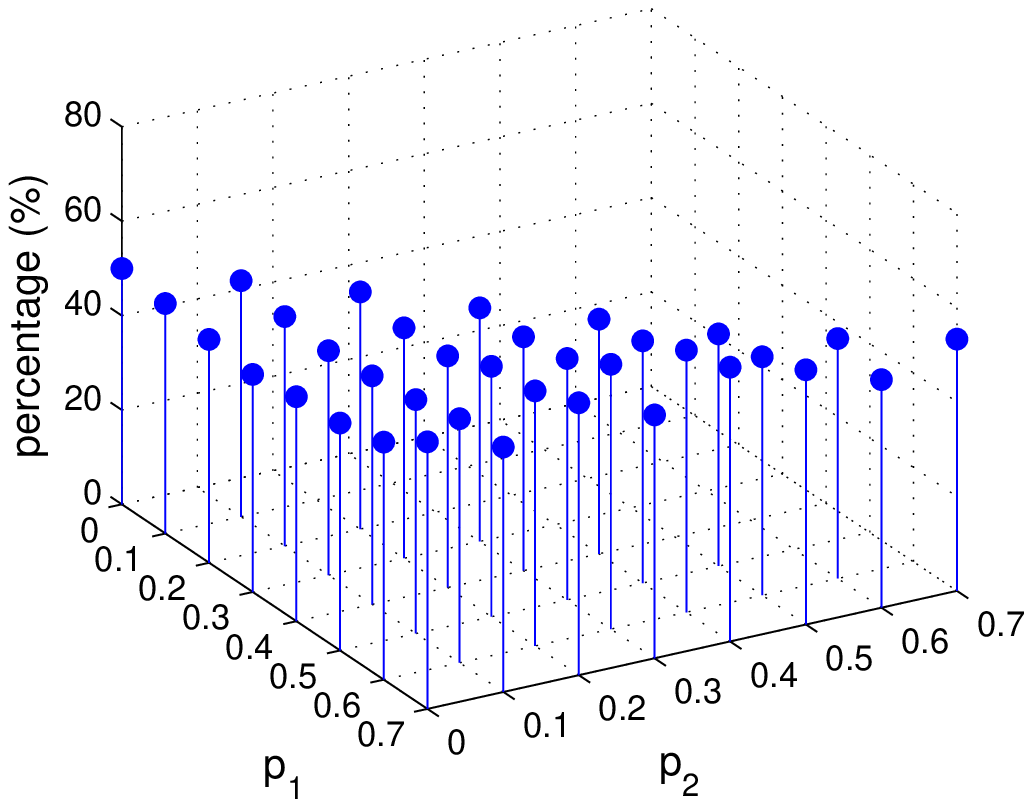}}
    \subfigure[CAFS]{
    \includegraphics[width=2in]{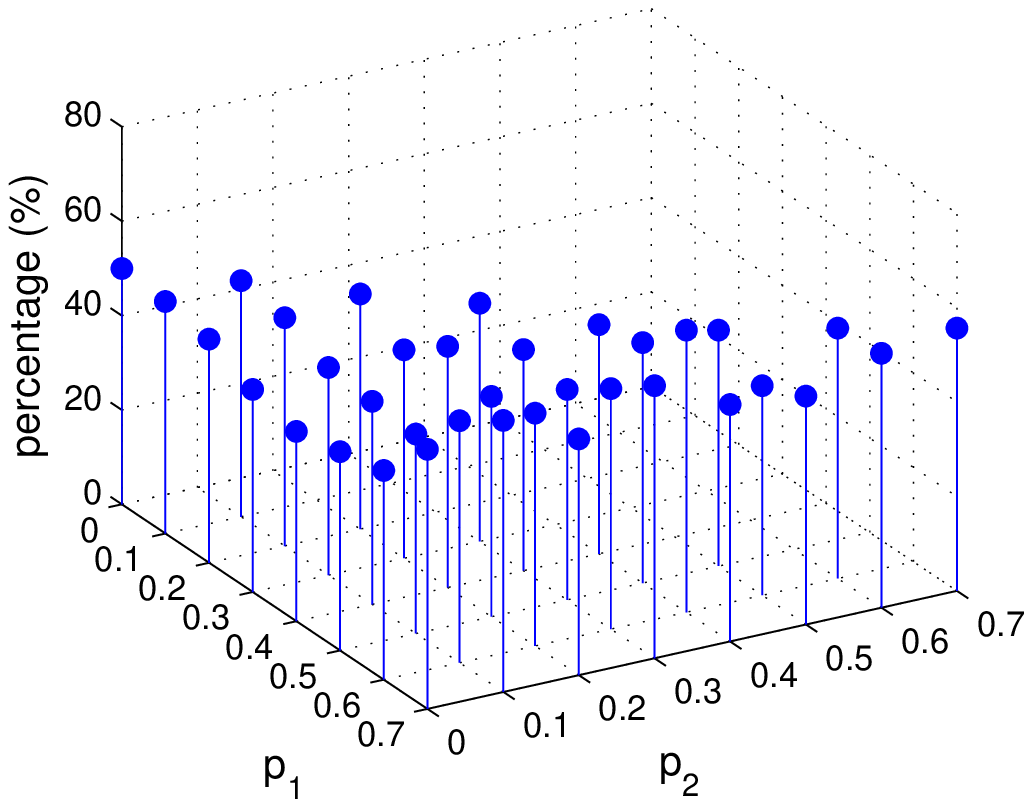}}
    \subfigure[VNFS]{
    \includegraphics[width=2in]{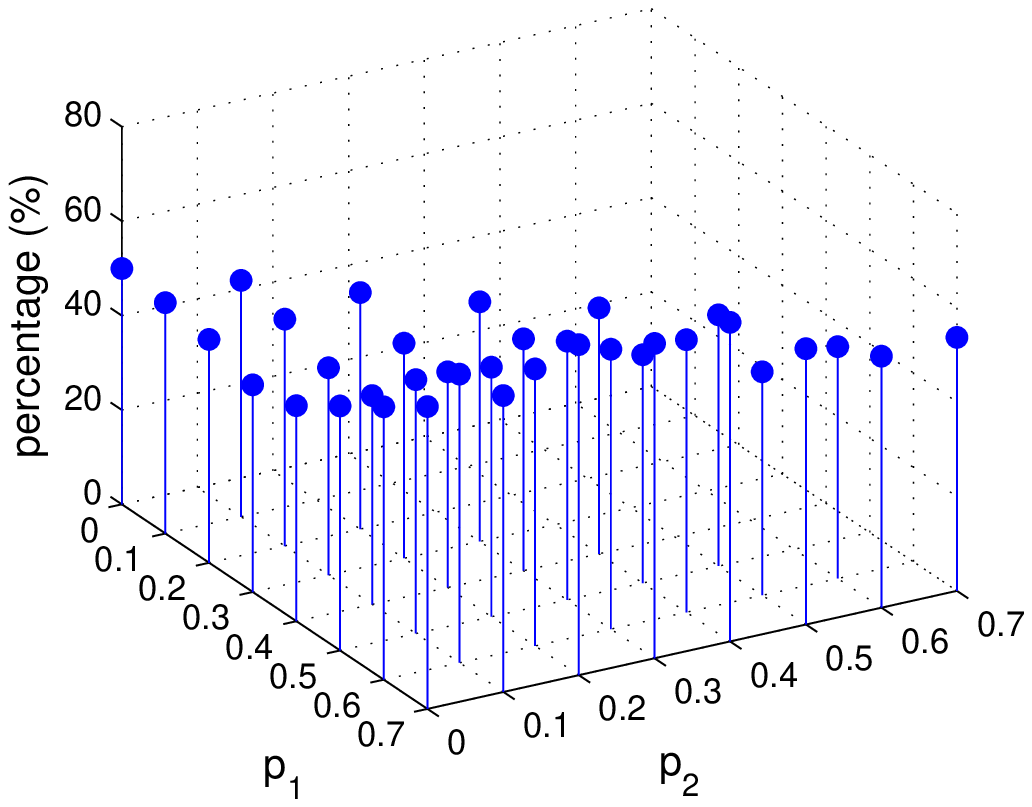}}
    \subfigure[VLFS]{
    \includegraphics[width=2in]{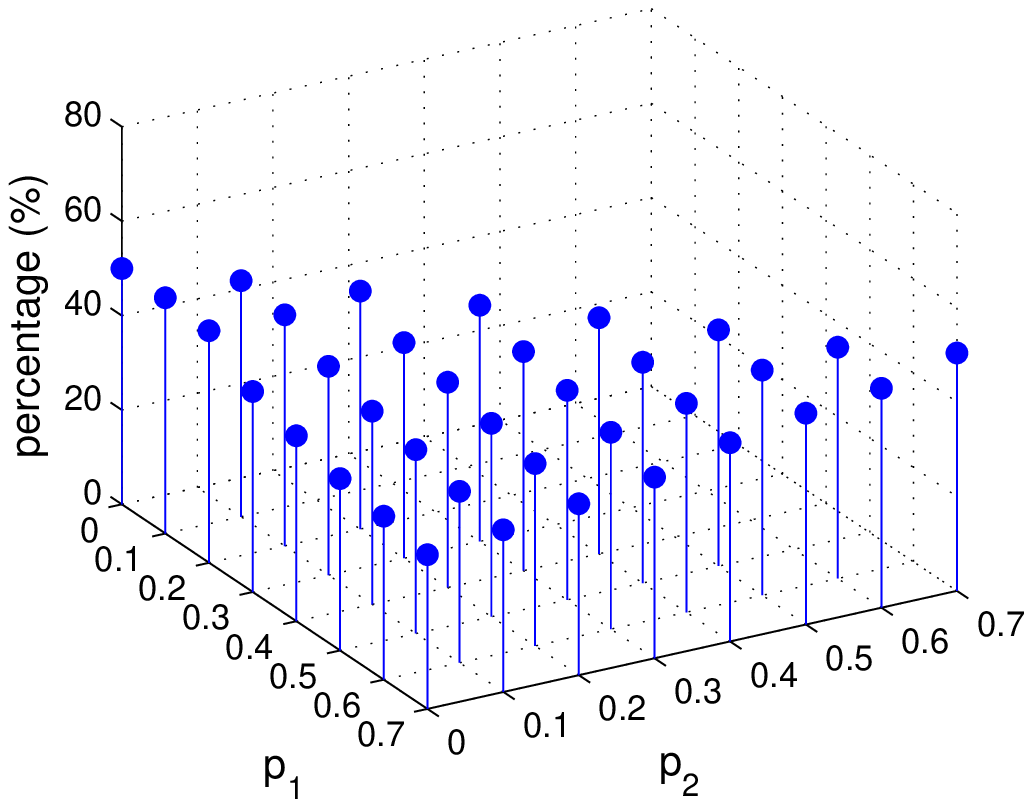}}
    \caption{Mean percentages of vehicles diverting on route 1 with different combinations of $p_1$ and $p_2$.}
    \label{fig:AllSplitRate}
\end{figure}

To enhance in the direction, we present mean percentages
with different combinations of $p_1$ and $p_2$ in Figure \ref{fig:AllSplitRate}.
Clearly, all cluster-based strategies perform similarly in all given combinations, and TTFS, VNFS and VLFS also fail to reflect the difference in the asymmetric network.
It strengthens that the strategies are unable to reflect the difference of the slowdown probability in the more general asymmetric network.

We lastly revisit the success of the cluster-based strategies in the symmetric network.
These strategies work based on measuring the congestion cluster on alternative routes.
Although the slowdown probability is involved in the NS model,
the number, sizes, locations, etc. of the congestion clusters are basically approximate on the same routes according to the law of large numbers as the routes are quite long (2000 cells);
it is just like repeating trials with one probability, and the mean obviously tends to one value when the number of repetition is high.
Although various weights are introduced by different strategies, the generated coefficients on the same routes are still close.
%
Therefore, the strategies direct about half of vehicles on route 1 and the other half on route 2 as shown in Figure \ref{fig:SymmetricSplitRate};
it is similar with randomly or alternatively diverting.
Whereas for a symmetric network, it is obvious that a simple and effective way to achieve balance is randomly or alternatively diverting vehicles.
This is the reason that the strategies perform well in the symmetric network.

\section{Conclusions}
The paper evaluates the following eight prevalent route guidance strategies in an asymmetric traffic system with different slowdown behaviors on two alternative routes:
travel time feedback strategy, mean velocity feedback strategy, congestion coefficient feedback strategy, prediction feedback strategy, weighted congestion coefficient feedback strategy, corresponding angle feedback strategy, vehicle number feedback strategy, vacancy length feedback strategy.

The results show that only mean velocity feedback strategy is able to approximate user optimality in the more general asymmetric traffic system, although all strategies except (for travel time feedback strategy) can also make it in the symmetric system.
The causes are analyzed, and we finally suggest mean velocity feedback strategy if the authority intends to achieve user optimality in practical traffic systems.

\section*{Acknowledgements}
The research has been funded by 973 Program (2012CB725403),
the National Science Foundation of China (71131001)
and the Fundamental Research Funds for the Central Universities (2012JBM064).

\bibliographystyle{model2-names}
\bibliography{library}







\end{document}